\documentclass{nature_arxiv}

\title{Topological Superconductivity in a Phase-Controlled Josephson Junction}

\author{Hechen Ren$^{1}$, Falko Pientka$^1$, Sean Hart$^{1,3}$, Andrew Pierce$^1$, Michael Kosowsky$^1$, Lukas Lunczer$^2$, Raimund Schlereth$^2$, Benedikt Scharf$^4$, Ewelina M. Hankiewicz$^4$, Laurens W. Molenkamp$^2$, Bertrand I. Halperin$^1$, \& Amir Yacoby$^1$\footnote{e-mail: yacoby@g.harvard.edu}}

\begin{document}

\maketitle
\begin{affiliations}
 \item Department of Physics, Harvard University, Cambridge, Massachusetts 02138, USA.
 \item Physikalisches Institut (EP3), Universit\"at W\"urzburg 97074, W\"urzburg, Germany.
 \item IBM T. J. Watson Research Center, Yorktown Heights, New York 10598, USA.
 \item Institute  for  Theoretical  Physics  and  Astrophysics, University  of  W\"urzburg,  97074  W\"urzburg,  Germany.
\end{affiliations}

\begin{abstract}
Topological superconductors can support localized Majorana states at their boundaries. These quasi-particle excitations have non-Abelian statistics that can be used to encode and manipulate quantum information in a topologically protected manner. While signatures of Majorana bound states have been observed in one-dimensional systems, there is an ongoing effort to find alternative platforms that do not require fine-tuning of parameters and can be easily scalable to large numbers of states. Here we present a novel experimental approach towards a two-dimensional architecture. Using a Josephson junction made of HgTe quantum well coupled to thin-film aluminum, we are able to tune between a trivial and a topological superconducting state by controlling the phase difference $\phi$ across the junction and applying an in-plane magnetic field. We determine the topological state of the induced superconductor by measuring the tunneling conductance at the edge of the junction. At low magnetic fields, we observe a minimum in the tunneling spectra near zero bias, consistent with a trivial superconductor. However, as the magnetic field increases, the tunneling conductance develops a zero-bias peak which persists over a range of $\phi$ that expands systematically with increasing magnetic fields. Our observations are consistent with theoretical predictions for this system and with full quantum mechanical numerical simulations performed on model systems with similar dimensions and parameters.  Our work establishes this system as a promising platform for realizing topological superconductivity and for creating and manipulating Majorana modes and will therefore open new avenues for probing topological superconducting phases in two-dimensional systems.

\end{abstract}

Majorana bound states (MBS) are quasiparticle excitations that are their own antiparticles and hence can serve as the basis of topological quantum computing, where quantum information can be stored and manipulated robustly\cite{
Majorana1937,Kitaev2000, Kitaev2003,Lutchyn2010, Oreg2010,Nayak2008,Beenakker2011,Stanescu2011,Alicea2012}. Spectroscopic studies have been conducted in various one-dimensional systems such as proximitized nanowires and atomic chains, where zero-bias peaks exist in the tunneling spectroscopy in individual parts of the parameter space associated with MBS\cite{
Mourik2012a, Rokhinson2012, Churchill2013, Das2012, Finck2013, Albrecht2016, Chen2017, Deng2016, Gul2018, Zhang2017, Nadj-Perge2014, Li2016}. Despite the growing evidence, scalable networks of Majorana qubits have proven a challenge to obtain in such one-dimensional platforms, due to both the intrinsic instabilities associated with one-dimensional systems and the technological obstacles in their physical implementation\cite{
Shabani2016,Karzig2017a}. Therefore, to understand and harvest the full power of MBS physics, two-dimensional platforms that can realize topological superconductivity are in demand for patterning large-scale networks of Majorana devices as well as integrating them with other quantum information devices and systems, in a reproducible and controlled fashion. 

We design and implement a controllable two-dimensional platform for realizing topological superconductivity based on a recent theoretical proposal for a planar Josephson junction created out of a two-dimensional electron gas (2DEG) subject to a strong Rashba spin-orbit interaction, sandwiched between two aluminum superconducting leads (Figure 1a)\cite{
Pientka2017}. In this system, the phase transition between trivial and topological superconductivity can be tuned using two independent knobs - the phase difference across the junction $\phi$, and the Zeeman energy $E_Z$ governed by an external magnetic field applied in the plane of the junction. In a long Josephson junction which is translationally invariant along $x$, the direction parallel to the superconducting electrodes, supercurrent is carried by bands of Andreev bound states in the normal region, formed by successive Andreev reflections at the normal-superconductor interfaces\cite{Andreev1964,Beenakker1991}. The energy of each Andreev state thus depends both on $\phi$ and on the phase accumulated by the quasiparticles traversing the junction at various angles from the $x$-direction. Therefore, the Andreev states can have a full range of wavevectors, with the $x$-components of their wavenumbers $k_x$ varying in magnitude from $0$ to nearly the Fermi wavenumber $k_F$. Their energies hence disperse to form a continuous sub-gap spectrum. Interestingly, when normal reflection at the normal-superconductor interface is weak, these Andreev bands are relatively flat and disperse only weakly with changing $k_x$ (Figure 1e). This leads to a strongly enhanced density of states near zero energy in the vicinity of the topological phase transition. When normal reflection is taken into account, the bands acquire a nonzero width (Figure 1f).

A topological phase transition in a clean junction is accompanied by a zero-energy crossing at $k_x = 0$ (Figure 1b). When the Zeeman energy is zero, the $k_x = 0$ Andreev states are twofold degenerate and cross at $\phi = \pi$ in the absence of normal reflections. A finite magnetic field parallel to the x-axis separates the $k_x =0$ states by a phase difference $\Delta \phi = 2 \pi E_Z / E_T$, where $E_Z$ is the Zeeman energy and $E_T = ( \pi/2) (\hbar v_F/W)$ is the Thouless energy. In the range of $\phi$ between these two crossings, the occupancy of fermionic states becomes odd, and the system undergoes a phase transition into a topological superconducting phase. We can map out this phase boundary in the $\phi - E_Z$ space (Figure 1c). As $E_Z$ increases from $0$ to the Thouless energy $E_T$, the junction becomes topological in a growing range of $\phi$, with predicted MBS on the end of a semi-infinite junction. As $E_Z$ further increases beyond $E_T$, this range in $\phi$ starts to decrease, forming overall diamond shapes (dashed lines in Figure 1c). In a physical system, normal reflection can occur at the interfaces, which hybridize the left- and right- moving modes in the junction and shift the phase boundary from the ideal case. In this case, the topological phase occupies regions deformed from the diamond shapes, but its dependence on magnetic field and phase difference are robust (solid colour in Figure 1c) and largely insensitive to changes in geometry and electron chemical potential. Consequently, for a wide range of magnetic fields, the application of a small phase bias can easily tune the system in and out of the topological superconducting phase and is hence a powerful experimental knob that can be controlled in a rapid manner.

Our planar Josephson junction consists of an 8-nm-wide HgTe quantum well contacted by thermally evaporated aluminum leads about 15 nm thick, with 5  nm of as a sticking layer. The junction region is 600 nm wide and about 1 micron long, with one end of each lead connected to form a flux loop. As previously established, our HgTe quantum well, grown by molecular beam epitaxy, provides a 2DEG with high mobility and dominant Rashba spin-orbit coupling, and the thin aluminum leads can superconduct up to 1.8 T of in-plane magnetic fields\cite{
Hart2014,Hart2017}. Using our vector magnet, we apply a magnetic field $B_z$ perpendicular to the sample plane to generate the flux that controls the phase difference across the junction, while we apply an in-plane magnetic field $B_x$ in the $x$-direction to tune the Zeeman energy (Figures 1a and 1d). On the outer edge of the junction, we fabricate a weakly coupled electrode by evaporating 10 nm of titanium and 100 nm of gold, with a few nanometers of CdHgTe as a tunnel barrier. The overlapping area is approximately 100 nm by 100 nm, which gives a tunneling resistance of around 300 k$\Omega$.

By applying an AC excitation in addition to a DC voltage bias on the tunnel probe and measuring the AC current through the superconducting lead, we can obtain a two-terminal differential conductance curve as we vary the DC bias. The observed spectrum of the proximitized 2DEG typically exhibits two broad coherence peaks separated by about 120 $\mu$V in bias voltage and a valley near zero bias (Figure 2a). To see how the spectrum disperses with the phase difference $\phi$, we scan $B_z$ over a few mT near zero and record the differential conductance as a function of both the bias voltage and $B_z$. Shown as colourmaps in Figure 2a-c, the tunneling spectrum exhibits a strong periodic modulation with $B_z$, where the period matches the area of the flux loop, considering the magnetic flux repelled by the superconducting lead. The in-plane field also generates an asymmetry between positive and negative voltage biases, which we attribute to a particle-hole asymmetry. To highlight the contribution to the tunneling conductance near zero energy we symmetrize the data at positive and negative biases (the raw data and further discussion are presented in the Supplementary Section 3).  

At low in-plane fields, the tunneling spectra reveal a conductance minimum near zero bias irrespective of the applied phase difference across the junction. This behaviour (Figures 2a and 2b) resembles recent measurements in a graphene Josephson junction and is interpreted as the behavior of the bulk Andreev bound states and their dependence on the phase difference across the junction\cite{Bretheau2017}. We attribute the missing zero-bias conductance peak at low in-plane fields near a phase difference of $\pi$ to the presence of weak normal reflections at the normal-superconducting interface. At high in-plane fields, a conductance peak develops near zero-bias over a range of $\phi$, repeating periodically (Figures 2c and 2f). The emergence of a robust and extended zero-bias peak in $\phi$ indicates the spectrum of the sub-gap states concentrating near zero energy and persisting over a wider span in $\phi$ as the in-plane magnetic field increases (Figures 3a-h).

To fully capture how the range in phase containing the zero-bias peak (ZBP) grows with magnetic field, we quantify the emergence of this conductance peak by extracting the curvature of the differential conductance curve around zero-bias, using a parabolic fit on the symmetrized data. We perform this analysis at all values of phase difference $\phi$ and in-plane field $B_x$ to produce a colourmap visualizing the development of the zero-bias peak in the $\phi - B_x$ phase space (Figure 3i). At low fields, most values of the phase difference give a dip (positive curvature) in the zero-bias conductance, shown in red in Figure 3i, indicative of a conventional superconducting phase. As $B_x$ increases, the differential conductance becomes flatter near zero bias, and the parabolic fit yields a small absolute value, giving rise to a white region in the colour plot, which expands to occupy higher fractions of each period in $\phi$. As the magnetic field continues from 0.6 T to 1.2 T, a blue region of negative curvature emerges, marking the zero-bias peak, and expands to fill the entire period. Similar behaviour of expanding ZBP region repeats for negative values of $B_x$, generating a phase diagram consistent with the predicted topological phase transition (Figure 1c). 

To simulate transport through the device, we describe the semiconductor by a tight-binding model with uniform Rashba spin-orbit coupling defined in a rectangular region. The region comprises a normal part sandwiched between two segments with proximity-induced superconductivity held at different phases (Supplementary Figure S14, similar to Figure 1a). We evaluate the conductance between a metallic tunneling probe attached to the edge of the normal region and two grounded superconducting leads that contact the superconducting regions on both sides, employing a scattering matrix approach using the KWANT package\cite{
Groth2014}.

Motivated by the experimental observation that the superconducting coherence peaks do not shift in energy with the in-plane magnetic field, we assume a suppressed g-factor in the proximitized parts and neglect the Zeeman field outside the normal region. Our model includes doping of the semiconductor due to the superconductor by assuming a higher density in the proximitized parts of the semiconductor. We account for the experimental resolution and nonuniform phase differences across the junction due to flux focusing by artificially broadening the theoretical data in energy and phase.

The calculated conductance, plotted as a function of bias voltage and phase in Figure 4a-f, reproduces key features of the experimental data. At low Zeeman fields, the heights of the coherence peaks at $V \sim \pm 70 \mu$V are modulated in phase and the conductance has a dip at zero bias for all values of the phase. At fields above $B_x = 0.5$ T, a peak at zero bias develops while the coherence peaks remain visible. In a minor deviation from the experiment, the coherence peaks shift to slightly higher energies $V \sim \pm 100 \mu$V in the numerics, which can be reconciled by accounting for a small gap suppression by the Zeeman effect in the superconductor.

The emergence of the zero-bias peak at finite fields is clearly visible in Figure 4g, where the curvature of the zero-bias conductance with bias voltage is plotted as a function of phase difference and the magnetic field. In the experimental field range, the numerical results in Figure 4g are in excellent agreement with the experimental data in Figure 3i. For all phase differences, the curvature monotonously decreases with in-plane field and eventually transitions from a dip to a peak. At $B_x = 1$ T, a zero-bias peak exists for all values of the phase.

Comparing the theoretical curvature plot in Figure 4g with the phase diagram in Figure 1c reveals that the most pronounced zero bias peaks occur close to the topological phase boundaries. At the same time, a zero-bias dip exists deep inside the topological phase. This is consistent with our numerical findings that the Majorana wavefunction is almost completely delocalized over the junction area for the experimental sample dimensions (see Supplementary Section 7), precluding a dominant Majorana signature in the measurement. Instead, the peak originates from a band of quasi-one-dimensional subgap states living inside the junction (Figure 1e). The density of states is enhanced at low energies as the band crosses zero energy in the vicinity of the topological phase transition, which manifests itself as a zero-bias conductance peak when the energy broadening is larger than the induced gap in the quasi-one-dimensional band. In the presence of normal reflection, the band acquires a finite width (Figure 1f), and the zero-bias conductance peak can exist in a broader parameter window around the phase transition (see Supplementary Section 8 for a discussion of the density of states).

In conclusion, our experiment provides the first tunneling spectroscopy study of phase-controlled Josephson junction in the presence of a tunable in-plane magnetic field. The ability to phase-bias the Josephson junction offers a powerful knob with easy access and control. Our measurements of a ZBP that develops and expands with applied magnetic field provide evidence of a topological phase transition in a two-dimensional induced superconductor. Our experiment can be easily generalized to other two-dimensional materials, where the interplay of phase bias, spin-orbit coupling, and Zeeman effect can create exciting opportunities to investigate topological superconductivity, making such platforms promising candidates for the detection and manipulation of Majorana bound states, and hence for realizing topologically protected quantum computation.

\begin{addendum}
 \item This work is supported by the NSF DMR-1708688, by the STC Center for Integrated Quantum Materials under NSF Grant No. DMR-1231319, and by the NSF GRFP under Grant DGE1144152. This work is also partly supported by the Army Research Office and was accomplished under Grant Number W911NF-18-1-0316. The views and conclusions contained in this document are those of the authors and should not be interpreted as representing the official policies, either expressed or implied, of the Army Research Office or the U.S. Government. The U.S. Government is authorized to reproduce and distribute reprints for Government purposes notwithstanding any copyright notation herein. We acknowledge additional financial support from the German Research Foundation (The Leibniz Program and Sonderforschungsbereich 1170 Tocotronics), the EU ERC-AG program (Project 4-TOPS), and the Bavarian Ministry of Education, Science and the Arts (IDK Topologische Isolatoren and the ITI research initiative). E.M.H. and B.S. acknowledge ﬁnancial support from the German Science Foundation (Leibniz Program, SFB1170 ”ToCoTronics”) and the Elitenetzwerk Bayern program “Topologische Isolatoren”.
 \item[Author Contributions]
The experiment is a collaboration between the Harvard and W\"urzburg experimental groups. H.R., F.P., B.S., E.M.H., B.I.H., and A.Y. carried out the theoretical modeling and analysis.
 \item[Competing Interests] The authors declare that they have no
competing financial interests.
\end{addendum}

\clearpage
\bibliography{references}
\addcontentsline{toc}{chapter}{References}  

\begin{figure}  
\centering
\includegraphics[width=0.96\textwidth]{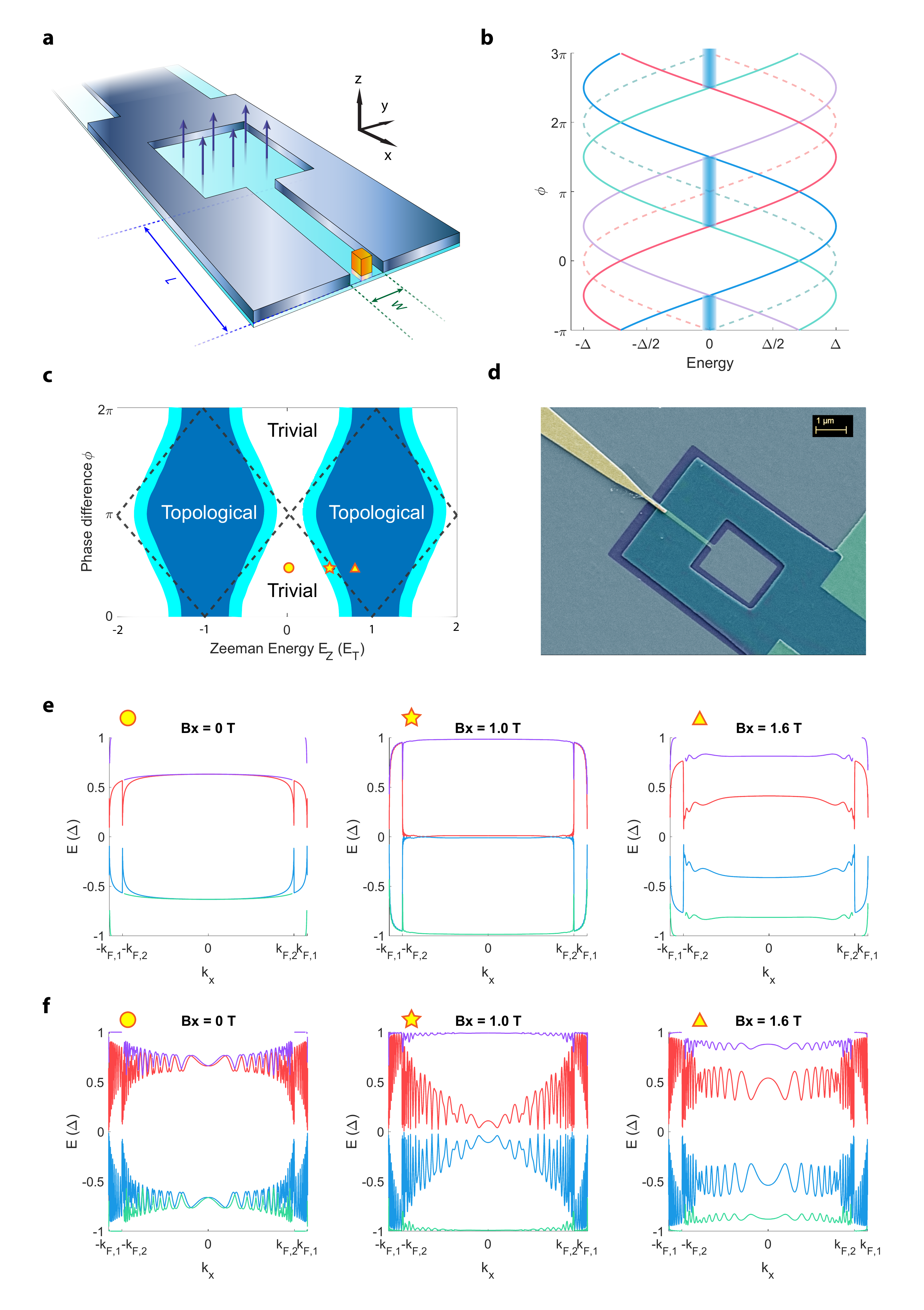}
\newpage
\caption[Topological Transition in a Phase-Controlled Josephson Junction]{\spacing{1.2}\textbf{Topological Transition in a Phase-Controlled Josephson Junction.} a, Device schematics for a planar Josephson junction created out of a 2DEG (shown in cyan) sandwiched between two aluminum superconducting leads, whose ends connect to form a flux loop (shown in steel blue)\footnote{We make devices both with and without the hole inside the flux loop and found little difference between the two designs. The data presented in the main text here is taken from a device with no hole in the middle.}. The tip of the tunnel probe (shown in gold) overlaps the HgTe quantum well, separated by a region of CdHgTe (shown in pastel colours). The perpendicular component of the external magnetic field $B_z$ is used to tune the phase difference across the junction, while its in-plane component $B_x$, parallel to the superconducting-normal interface, is used to tune the Zeeman energy. b, The bound-state spectrum for $k_x =0$ for a junction that is long in the $x$-direction and symmetric in the $y$-direction. It is twofold degenerate in the absence of any external magnetic field (dashed lines), and spin-split in the presence of an external in-plane field (solid lines in rose, blue, violet, and cyan for the spin-up and spin-down electrons and holes), accommodating a topological phase to develop in the range of $\phi$ between the zero-bias crossings (shaded regions coloured blue). c, Phase diagram as a function of the Zeeman energy $E_{Z}$ (given in units of the Thouless energy $E_T$) and the phase difference $\phi$ across the junction. The dashed lines correspond to the boundary between the topological and trivial phases of superconductivity for a junction with perfect transparency, while the solid blue regions show the topological phase can deviate from the diamond shapes when considering normal reflection at the superconducting-normal interface. The cyan ribbons highlight the boundary between the two phases of superconductivity. d, False-colour scanning electron micrograph of a device with a narrow junction. Green region defines the mesa area which contains the HgTe quantum well. The superconducting contact is Ti/Al (5 nm/15 nm) and coloured in purple. The tunnel probe is Ti/Au (10 nm/100 nm) and coloured in yellow. Data presented in the main text is taken from a device with a wider junction (600 nm) and without the hole in the mesa. e, Dispersion of the Andreev band as a function of $k_x$, at three different values of the Zeeman energy. The circle, the star, and the triangle symbols indicate where each band diagram corresponds in the phase diagram in 1c. The Andreev bands become relatively flat during the topological phase transition (middle panel). f, Similar Andreev spectra as calculated in e but including some normal reflection, which results in a finite width of the bands.

\label{fig:fig1}}
\end{figure} 

\newpage

\begin{figure} 
\centering
\vspace*{20 mm}
\includegraphics[width=\textwidth]{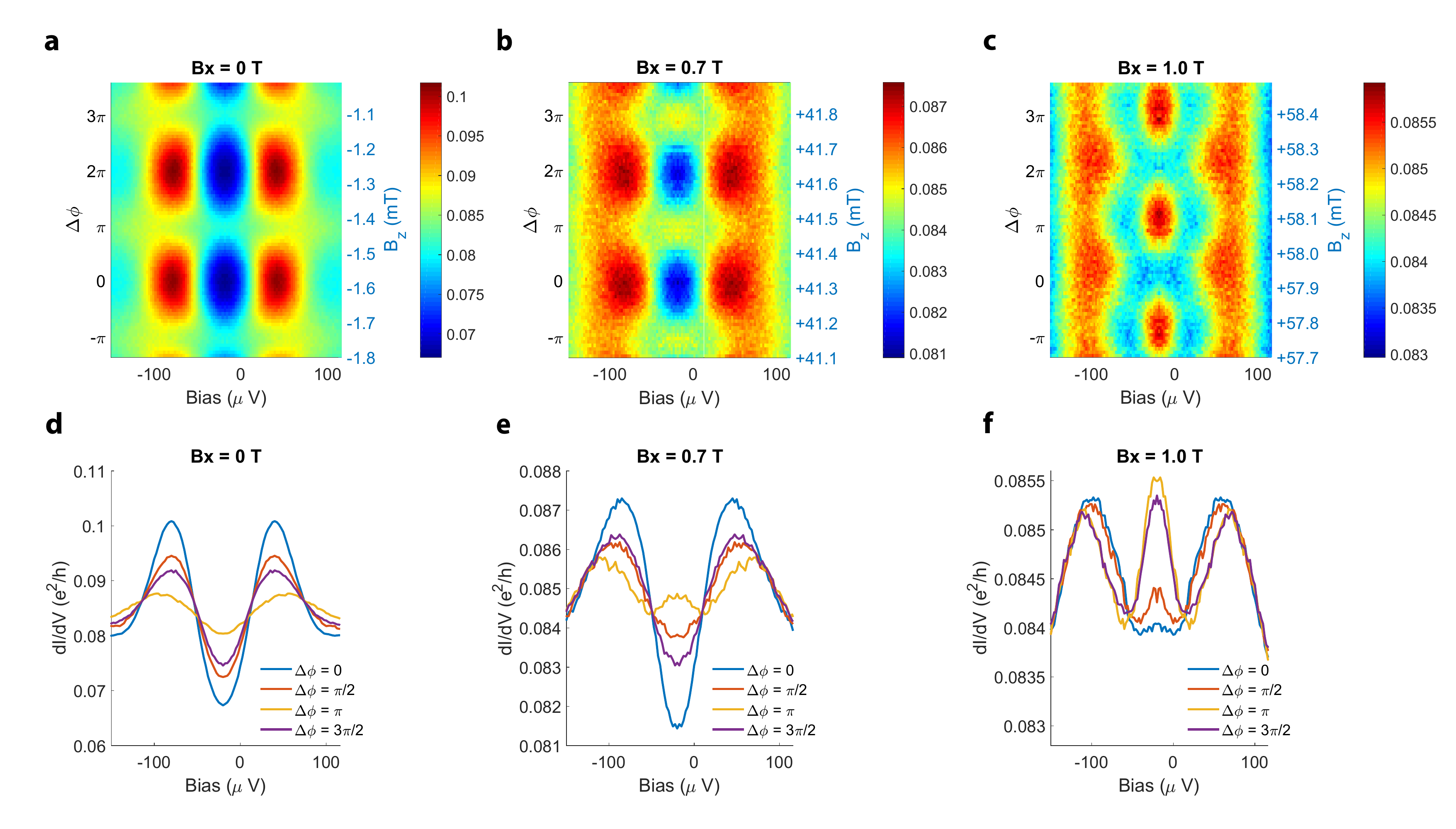}
\caption[Phase modulations of the tunneling conductance at low versus high fields]{\spacing{1.2}\textbf{Phase modulations of the tunneling conductance at low versus high fields.} a-c, Differential conductance colour plots (in units of $e^2/h$) as a function of both the bias voltage ($x$-axis) and the phase difference (left $y$-axis) offset from the $\Delta \phi = 0$ point, identified by the value of $\phi$ at which the coherence peaks obtain a maximum. The right $y$-axis records the actual $B_z$ field supplied by the magnet. d-f, Linecuts of a-c showing the differential conductance curves as a function of the bias voltage on the tunnel probe at four representative values of phase differences, averaged over repeating lines spaced integer periods apart. a and d present data taken at $B_x = 0$ T, where no zero-bias peak exists for any phase difference. b and e present data taken at $B_x = 0.7$ T, where a zero-bias peak exists for a range of phase differences within each period. c and f present data taken at $B_x = 1.0$ T, where a zero-bias peak persists through most of the period.  
\label{fig:fig2}}
\end{figure} 

\newpage
\begin{figure} 
  \centering
\includegraphics[width=\textwidth]{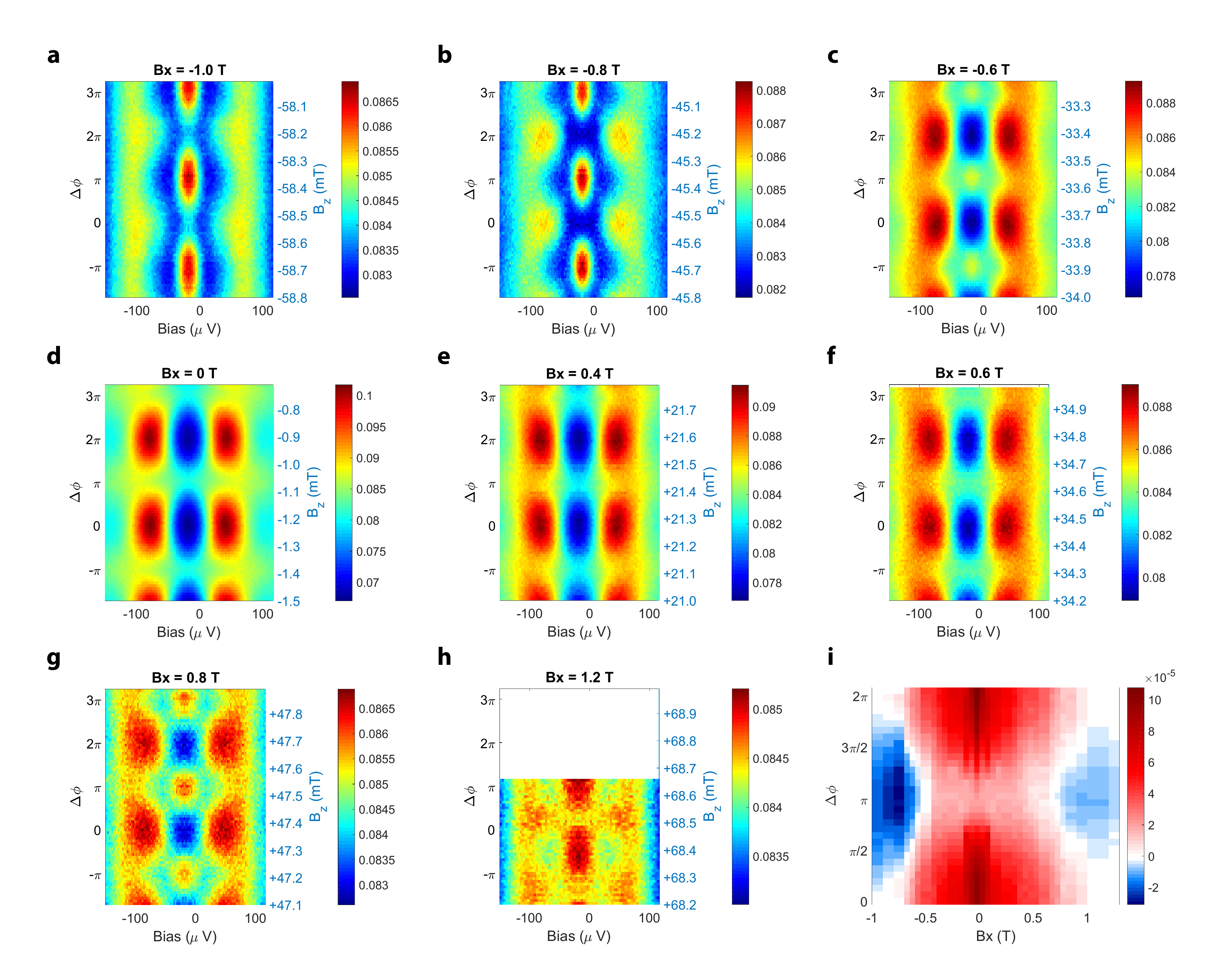}
\caption[Development of zero-bias peak and the reconstructed phase diagram]{\spacing{1.2}\textbf{Development of zero-bias peak and the reconstructed phase diagram.} a-h, Progression of tunneling conductance colourplots (in units of $e^2/h$) as the magnetic field $B_x$ varies from $-1.0$ T to $1.2$ T, skipping field values shown in Figure 2 except for $B_x = 0$ T which serves as a point of reference. All plots span over a range of $0.8$ mT in $B_z$ except $ B_x = 1.2$ T. i, colourplot showing the extracted zero-bias curvature as a function of both the in-plane magnetic field $B_x$and phase difference offset $\Delta\phi$, shown in units of $ e^2/h \cdot \mu \text{eV}^{-2}$. The blue region in the phase diagram shows where a well-defined zero-bias peak is present in the tunneling conductance. Its emergence and expansion with the application of both positive and negative in-plane magnetic fields agrees with the predicted phase transition (Figure 1c).
\label{fig:fig3}}
\end{figure}

\newpage

\begin{figure} 
 \centering
\includegraphics[width=0.9\textwidth]{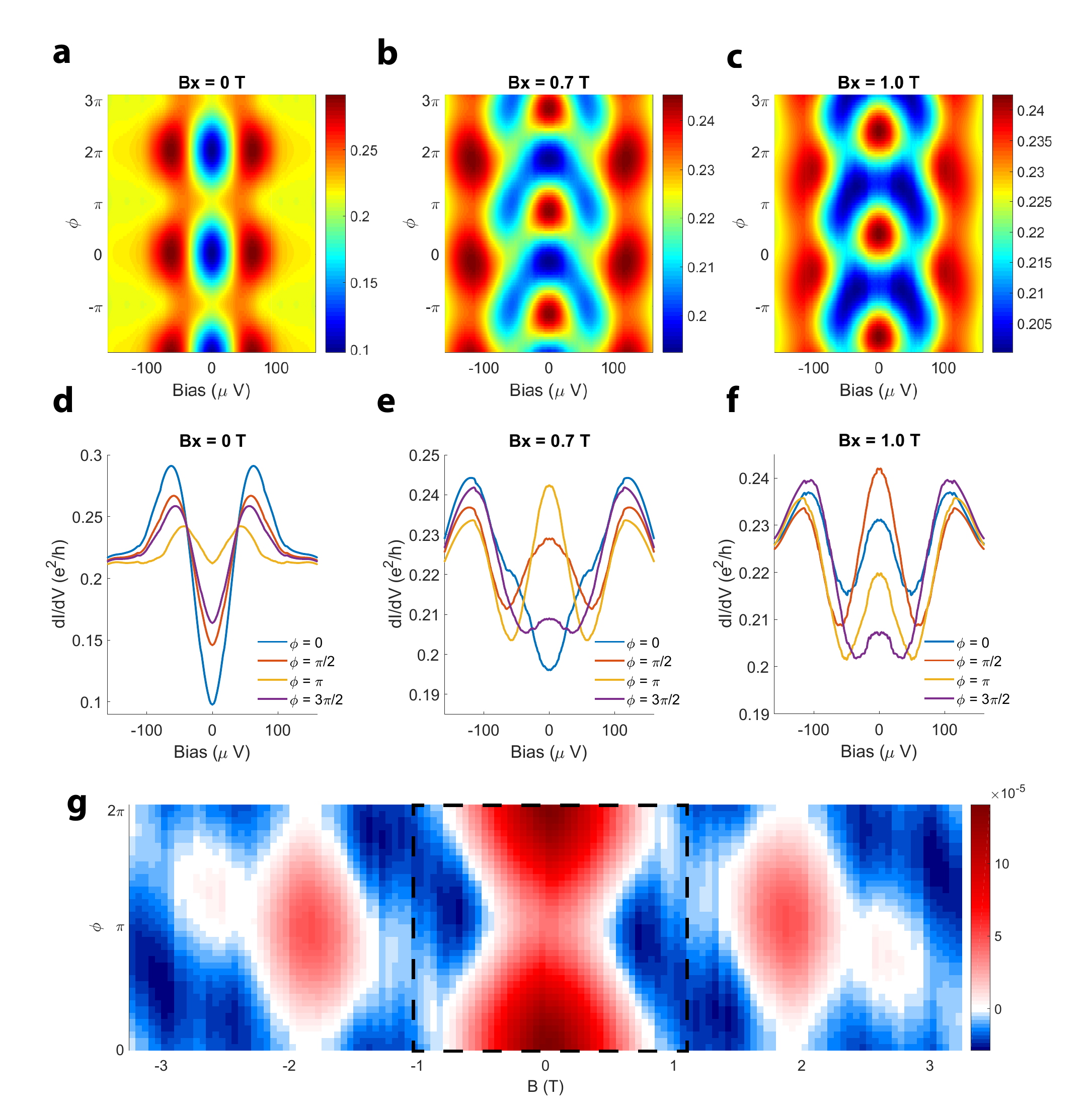}
\vspace*{-20 mm}
\caption[Numeric simulations for the tunneling conductance using a tight-binding model]{\spacing{1.2}\textbf{Numeric Simulation for the tunneling conductance using a tight-binding model.} a-f, Calculated tunneling conductance between a metallic tunneling probe attached to the edge of the normal region and two grounded superconducting leads. d-f are linecuts at four representative values of phase differences, taken from the conductance colour plots a-c, at three $B_x$ fields corresponding to those shown in Figure 2. The emergence of the zero-bias peak in finite fields agrees with the experimental data. g. Predicted zero-bias curvature of the tunneling conductance as a function of both the Zeeman energy $E_Z$ and the phase difference $\phi$, shown in units of $ e^2/h \cdot \mu \text{eV}^{-2}$. Indicative of a zero-bias peak, the blue regions trace out the transition between the trivial and topological superconducting phases. Outlined in black dashed lines is the regime corresponding to the experimental data.
\label{fig:fig4}}
\end{figure}



\section*{Supplementary material}

# Supplementary Information for Topological Superconductivity in a Phase-Controlled Josephson Junction

Hechen Ren[1], Falko Pientka[1], Sean Hart[1,3], Andrew Pierce[1], Michael Kosowsky[1], Lukas Lunczer[2], Raimund Schlereth[2], Benedikt Scharf[4], Ewelina M. Hankiewicz[4], Laurens W. Molenkamp[2], Bertrand I. Halperin[1], & Amir Yacoby[1]*

[1] *Department of Physics, Harvard University, Cambridge, Massachusetts 02138, USA.*

[2] *Physikalisches Institut (EP3), Universität Würzburg 97074, Würzburg, Germany.*

[3] *IBM T. J. Watson Research Center, Yorktown Heights, New York 10598, USA.*

[4] *Institute for Theoretical Physics and Astrophysics, University of Würzburg, 97074 Würzburg, Germany.*

## 1  Device Characterization and Measurement

To characterize our two-dimensional electron gas (2DEG), we fabricate and measure a van der Pauw device on the same HgTe wafer QC0261, using the same metal deposition method described in the main text except replacing the aluminum contacts with gold. By measuring the longitudinal and Hall resistances, we obtain a 2DEG mobility of around 400,000 cm$^2$/Vs at a density of $1.0 \times 10^{11}$ cm$^{-2}$. The longitudinal and Hall resistances are shown respectively in Figure S1a and b.

To investigate the spectrum at the end of the Josephson junction, we perform local tunneling

---

*e-mail: yacoby@g.harvard.edu

spectroscopy through a weakly coupled electrode, which overlaps the junction over an area of approximately 100 nm by 100 nm near its edge. We use in-situ ion milling to etch most of the CdHgTe barrier layer on top of the quantum well and stop when only a few nanometers of CdHgTe remain. We then deposit 100 nm of gold with 10 nm of titanium as a sticking layer. Both metals, especially the titanium, react chemically with CdHgTe to form alloys[1]. These semimetallic alloys facilitate the tunneling of electrons between the local electrode and the quantum well, producing low-temperature tunnel resistances ranging from 15 k$\Omega$ to 15 M$\Omega$, depending on the thickness of the barrier layer and the overlapping area of the electrode with the quantum well.

Since the features in the tunneling conductance we study here are induced by the superconductivity in the aluminum, we expect them to only develop well below the critical temperature of the superconducting aluminum film, which can range from 1.2K to 1.6K [2]. Figure S3 shows the temperature dependence of the tunneling conductance curves during a cool-down process in our dilution refrigerator, for the same device discussed in the main text. As we can see, the main peaks in the conductance only start developing below 0.5 Kelvin. The separation of the two conductance peaks is typically around 120 $\mu$eV and the width of each peak is on the order of 50 $\mu$eV in energy. Above this temperature, the tunneling conductance curve is mostly flat for this energy range. Although this does not completely rule out all possible contributions from features related to the tunnel probe that only develop at low temperatures, it does strongly suggest that these conductance peaks originate from the aluminum-induced superconducting gap in the HgTe quantum well.

During our measurements, the point in bias voltage about which the conductance curve is

2

approximately symmetric is constant for each cool-down of the device, and we identify this point as the zero-bias point. It is worth noticing that this true zero-bias point tends to be offset by 15-25 $\mu$eV from zero reading on our DC voltage supply, which we attribute to a systematic offset caused by instrumental errors. Because it remains constant throughout our measurements, this offset causes no serious effect on the physics we observe, and hence we keep the x-axis reading unprocessed.

## 2 Correction for Imperfect Sample-Magnet Alignment

Due to imperfect sample-magnet alignment, the in-plane field $B_x$ will generate a small perpendicular component $B_z$, which offsets the $B_z$ field corresponding to zero-flux in the phase loop. To correct for this effect and maintain our measurement in a range of phase difference near zero, we scan over a range of $B_z$ at each $B_x$ field and identify a range with maximal oscillation amplitude in the tunneling conductance.

Shown in Figure S2a is the conductance in the range of a few mT in $B_z$ near zero, featuring periodic oscillations as the junction's phase difference is modulated, which gradually dampens as the $B_z$ increases in both positive and negative directions. Based on this experimental observation, we maintain our measurements in the vicinity of zero phase difference by scanning over a range in $B_z$ at each $B_x$ field and selecting a region with maximal oscillations in the tunneling conductance (Figure S2a and b). As the in-plane field $B_x$ increases, this optimal range drifts towards higher $B_z$ values, following a linear dependence to first order (Figures S2c-f).

The phase smearing effect is yet another high-field effect that is not compensated by this correction. Because of the microscopic roughness along the edge of the superconducting electrodes during the sample fabrication, the high in-plane magnetic fields can generate non-uniformity along the normal-superconducting interface, which in turn causes the phase difference along the $x$-direction of the junction to disperse[3]. This creates a smearing effect on the observed spectroscopy's dependence on phase, reducing the signal-to-noise ratio of the tunneling conductance. Because this effect worsens at high in-plane fields, it is necessary to adopt a junction geometry where the topological phase transition happens at low enough fields before the phase smearing effect masks the experimental signal entirely. In our experiment, the 600nm junction places us in this situation and enables us to see the emergence of the zero-bias peaks associated with the phase transition. Future development to improve the lithography and film quality of the superconducting contacts could help reduce such phase smearing effect and enhance the measurement resolution in both energy and phase at high in-plane fields.

## 3 The Electron-Hole Asymmetry with In-Plane Field

At zero in-plane field, we generally see a symmetric differential conductance between positive and negative DC voltages, endorsing the particle-hole symmetry in the local spectrum of Andreev bound states (see Main Text Figure 2a as an example). However, such symmetry in the LDOS is not present at high in-plane magnetic fields. This is reflected by an asymmetry in the differen-

4

tial conductance about zero bias voltage[1], which grows with the in-plane magnetic field applied. Figures S4 and S5 show the raw data with this asymmetry at various magnetic fields.

To visualize this asymmetry in the tunneling conductance and see how it develops with the magnetic field, we can average the differential conductance over an entire period in the phase difference, at each in-plane field, as plotted in Figure S7. Furthermore, we can quantify it with the slope of a linear fit to the averaged conductance curve. Shown in Figure S8, the slope representing the asymmetry grows linearly with the magnitude of the in-plane field, and the sign of the asymmetry does not depend on the direction of the field. This is also evident from Figure S6, which shows a continuous evolution as the in-plane field is ramped across a broad range, as well as a broader range in the bias voltage. The out-of-plane component of the $B_x$ field is passively compensated in linear proportion but not perfectly canceled. As a result of the minute remaining $B_z$ field, the zero-bias conductance shows oscillations in the $B_x$ field. We can see the same asymmetry develops from positive to negative $B_x$ fields, although its extent does vary from device to device.

The exact mechanism for controlling this asymmetry is not well understood and hence remains an interesting topic for further studies. Meanwhile, we can correct for this asymmetry, post-measurement, by averaging the differential conductances at each positive energy with its negative counterpoint and assigning this same value for both points. We perform this symmetrization procedure for each differential conductance trace at all phases and all fields, and the corrected results are shown in Figures 2 and 3 of the main text.

---

[1] As mentioned in Supplementary Section 1, this true zero bias can be offset from 0 voltage reading by a constant deviation ranging from 15 to 25 $\mu eV$, due to instrumental errors.

5

## 4  Obtaining the Zero-Bias Curvature

To illustrate the continuous opening of the window containing the zero-energy peaks, we analyze the symmetrized data and identify whether the differential conductance curve at each particular $B_z$ and $B_x$ contains a zero-energy peak. To quantify this property, we extract the local curvature of the differential conductance near zero-energy. For any differential conductance curve, we fit a parabola to its central segment, and hence extract the curvature from the fitting parameters. This is a good proxy for whether it contains a zero-energy peak as such peaks would make the center of the otherwise convex part of the curve concave, resulting in a negative curvature.

We obtain the curvature of the zero-energy conductance peak using parabolic fits. As seen in the fitting examples in Figure S9, the sign of the curvature correlates well with whether the curve contains a peak (panels e and f) or dip (panels a and c) at zero energy, and the amplitude roughly indicates its visibility. We can do this for each differential conductance curve at any particular phase difference $\phi$ and any in-plane field, $B_x$. The parabolic fit works the worst when the differential conductance is flat in the central region, as shown in panels b and d. In such cases, our method produces a rather small curvature, so the overall analysis is minimally affected. A colour map of the extracted curvature as a function of both phase difference $\phi$ and the Zeeman field $B_x$ is shown in Figure 3i of the main text.

## 5   Two Aadditional Devices with Similar Behavior

To demonstrate reproducibility for the general trend of the zero-bias curvature, here we present two other devices with different geometries but showing similar effects as the device shown in the main text. Device Kappa has a junction that is 400nm wide and 4 $\mu$m long, with a tunnel resistance around 12 M$\Omega$. The raw and symmetrized conductance data of Device Kappa are shown in Figures S10 and S11 respectively. Device Zeta has a junction that is 800nm wide and 1 $\mu$m long, with a tunnel resistance around 3 M$\Omega$. The raw and symmetrized conductance data of device Zeta are shown in Figures S12 and S13 respectively. Due to their low tunnel conductances, they give low signal-to-noise ratios, making them less immune to the phase smearing effects at high in-plane fields, as discussed in supplementary Section 2. Nevertheless, we see the same qualitative trend of an enhanced zero-energy conductance extending over a growing range of $\phi$ as the in-plane field increases.

We see again that the electron-hole asymmetry in the tunneling conductance is present in these two devices, and the slopes of this asymmetry in these two devices are opposite to that of the device discussed in the main text. Yet, just like in the case of the device shown in the main text, the sign of this slope is independent of the sign of the field. This suggests the asymmetry may be related to device-specific effects such as local potentials or disorders.

Moving towards narrower junctions brings potential benefits to the experiment. As we will show in Supplementary Section 10, a long junction which is 400nm wide holds the promise of showing zero-bias peaks both associated with the phase transition, in the field range we have mea-

sured, and from the Majorana bound states, at even higher fields. However, due to the phase smearing effect, the modulation of our tunneling conductance diminishes much below the field required to see the latter ZBP in this 400-nm-wide junction, and, therefore, we only observe the former type of peak in conductance. This points out an immediate direction for fabrication development to improve the lithographic definition and interfacial quality of the superconducting film with the quantum well, so that future experiments can go up to higher in-plane fields, and, therefore, down to narrower junctions. Such improvements will allow us to explore this topological phase transition physics and the associated Majorana bound states with a more discrete spectrum, without losing the experimental signal due to the undesired phase smearing effect.

## 6  Numerical Calculation of the Tunneling Conductance

Our theoretical description of the system is based on the model used in Reference 4 for a two-dimensional semiconductor with Rashba spin-orbit interaction defined on a rectangular region. The region has length $L$ and is divided into three segments: a normal region of width $W$ between two superconducting regions of width $W_{SC}$ (see rectangle with black contour in Figure S14). The Bogoliubov-de Gennes Hamiltonian can be written in the Nambu basis $(\psi_\uparrow, \psi_\downarrow, \psi_\downarrow^\dagger, -\psi_\uparrow^\dagger)$ as $H = \int d^2r \psi^\dagger(\mathbf{r}) \mathcal{H}(\mathbf{r}) \psi(\mathbf{r})$ with

$$\mathcal{H}(\mathbf{r}) = \left[\frac{p^2}{2m} - \mu(y) + \frac{m\alpha^2}{2}\right]\tau_z + E_Z(y)\sigma_x + \alpha(p_x\sigma_y - p_y\sigma_x)\tau_z + \Delta(y)\tau_+ + \Delta(y)\tau_-, \quad (1)$$

where $m$ is the effective mass, $\alpha$ the strength of the Rashba spin-orbit interaction and the $\sigma$ ($\tau$) are Pauli matrices in spin (particle-hole) space.

The Zeeman energy $E_Z(y) = g(y)\mu_B B/2$, chemical potential $\mu$ and pairing strength $\Delta$ are all affected by the proximity coupling between the semiconductor and the superconductor and therefore assume different values in the junction and in the leads. We write

$$\Delta(y) = \Delta[e^{-i\phi/2}\Theta(-y) + e^{i\phi/2}\Theta(y-W)], \tag{2}$$

$$g(y) = g_N\Theta(y)\Theta(W-y) + g_{SC}[1 - \Theta(y)\Theta(W-y)], \tag{3}$$

$$\mu(y) = \mu_N\Theta(y)\Theta(W-y) + \mu_{SC}[1 - \Theta(y)\Theta(W-y)], \tag{4}$$

where $\phi$ is the phase difference across the junction. Because of the experimental finding that the coherence peaks remain at the same energy for the entire magnetic field range, we henceforth assume the Zeeman energy to be completely suppressed in the superconducting part setting $g_{SC} = 0$ and denote $E_Z = g_N \mu_B B/2$.

For the numerical calculations it is convenient to consider the same model on a square lattice. The corresponding tight-binding Hamiltonian can be written in terms of the Nambu spinors $c_{i,j} = (c_{i,j,\uparrow}, c_{i,j,\downarrow}, c^\dagger_{i,j,\downarrow}, -c^\dagger_{i,j,\uparrow})^\mathrm{T}$ as

$$\begin{aligned} H = &-\sum_{i=1}^{L-1}\sum_{j=-W_{SC}+1}^{W+W_{SC}} [c^\dagger_{i,j}(t + i\alpha_{\mathrm{TB}}\sigma_y)\tau_z c_{i+1,j} + \mathrm{h.c.}] \\ &-\sum_{i=1}^{L}\sum_{j=-W_{SC}+1}^{W+W_{SC}-1} [c^\dagger_{i,j}(t - i\alpha_{\mathrm{TB}}\sigma_x)\tau_z c_{i,j+1} + \mathrm{h.c.}] \\ &+\sum_{i=1}^{L}\sum_{j=-W_{SC}+1}^{W+W_{SC}} \left[c^\dagger_{i,j}\left(4t + \frac{\alpha_{\mathrm{TB}}^2}{t} - \mu(j) + E_Z(j)\sigma_x\right)\tau_z c_{i,j} + (\Delta(j)c_{i,j}\tau_x c_{i,j} + \mathrm{h.c.})\right] \end{aligned} \tag{5}$$

In the limit $t \to \infty$, the tight-binding model maps onto the continuum model in Eq. (1) with the replacements $c_{i,j} \to a\psi(\mathbf{r})$, $\alpha_{\mathrm{TB}} \to \alpha/2a$ and $t \to 1/2ma^2$, where $a$ is the lattice constant.

9

For the numerical calculations presented in the main text, we use the following parameters in the continuum model of Eq. (1): $\alpha = 34\,\text{meVnm}$, $\Delta = 64\,\mu\text{eV}$, $\mu_N = 12.5\,\text{meV}$, $\mu_{SC} = 50\,\text{meV}$, $g_n = 10$, and $m = 0.033 m_e$, where $m_e$ is the electron mass.

The tight-binding parameters are chosen accordingly as $t = 32\,\text{meV}$, $\alpha_{\text{TB}} = 0.089 t$, and $a = 6\,\text{nm}$.

The phase diagram and subgap spectrum shown in Figure 1 of the main text is evaluated using the continuum model with $L = \infty$ and $W_{SC} = \infty$. In this case the momentum along the $x$ direction is conserved and can be replaced by a number $k_x$. The subgap spectrum can then be found by evaluating the scattering matrix $S(E)$ for states inside the junction and solving the equation

$$\det\{1 - S[E(k_x)]\} = 0 \tag{6}$$

(see Reference 4 for the numerical procedure). The phase boundaries correspond to zero-energy crossings at $k_x = 0$.

An evaluation of the conductance through the system requires the addition of external leads. To model the experimental setup, we attach a normal lead near the upper edge of the normal region and two superconducting leads contacting the lower edges of the superconducting segments as shown in Figure S14. We choose system dimensions $L = 150$, $W = 100$, and $W_{SC} = 200$ corresponding to a junction area of $900\,\text{nm} \times 600\,\text{nm}$.

The normal lead is added as a second layer, which extends infinitely in the $x \to -\infty$ direction and is connected to the junction by vertical tunnel couplings in the area $i_1 \leq i \leq i_2$ and $j_1 \leq j \leq$

10

$j_2$. The additional contribution to the Hamiltonian reads

$$H_L = \sum_{i=-\infty}^{i_2} \sum_{j=j_1}^{j_2} (4t - \mu_L) d^\dagger_{i,j} \tau_z d_{i,j}$$

$$- \left( t \sum_{i=-\infty}^{i_2-1} \sum_{j=j_1}^{j_2} d^\dagger_{i,j} \tau_z d_{i,j+1} + t \sum_{i=-\infty}^{i_2} \sum_{j=j_1}^{j_2} d^\dagger_{i-1,j} \tau_z d_{i,j} + t_L \sum_{i=i_1}^{i_2} \sum_{j=j_1}^{j_2} d^\dagger_{i,j} \tau_z c_{ij} + \text{h.c.} \right) \quad (7)$$

In our simulations, we choose $\mu_L = 4t$, $t_L = 0.2t$. The contact area is defined by $i_1 = 1$, $i_2 = 6$, $j_1 = 43$, $j_2 = 58$, which corresponds to an area of $\sim 40\,\text{nm} \times 100\,\text{nm}$.

The superconducting leads simply continue the two superconducting segments in the $x \to \infty$ direction

$$H_{SC,L} = -\sum_{i=L+1}^{\infty} \left( \sum_{j=-W_{SC}+1}^{0} + \sum_{j=W+1}^{W+W_{SC}} \right) \left[ \left( 4t + \frac{\alpha_{\text{TB}}^2}{t} - \mu_{SC} \right) c^\dagger_{i,j} \tau_z c_{i,j} + (\Delta(j) c_{i,j} \tau_x c_{i,j} + \text{h.c.}) \right]$$

$$- \sum_{i=L}^{\infty} \left( \sum_{j=-W_{SC}+1}^{0} + \sum_{j=W+1}^{W+W_{SC}} \right) [c^\dagger_{i,j}(t + i\alpha_{\text{TB}} \sigma_y) \tau_z c_{i+1,j} + \text{h.c.}]$$

$$- \sum_{i=L}^{\infty} \left( \sum_{j=-W_{SC}+1}^{-1} + \sum_{j=W+1}^{W+W_{SC}-1} \right) [c^\dagger_{i,j}(t - i\alpha_{\text{TB}} \sigma_x) \tau_z c_{i,j+1} + \text{h.c.}]. \quad (8)$$

We use the KWANT library to calculate the scattering matrix of the system with respect to the external leads. The conductance between the normal lead at a finite bias voltage $eV$ and the two grounded superconductors can then be evaluated according to the formula

$$G(eV) = \frac{e^2}{h}[N - R_{ee}(eV) + R_{eh}(eV)], \quad (9)$$

where $N$ is the number of channels in the lead and $R_{ee,eh}(\epsilon)$ are the normal and Andreev reflection probability in the normal lead at energy $\epsilon$. To account for the limited energy resolution in the

11

experiment, we convolute the conductance as a function of energy with a Gaussian with standard deviation $24\,\mu$eV. Moreover, we include a broadening of the conductance as a function of phase difference by convolution with a Gaussian with standard deviation $0.4\pi$. This accounts for an inhomogeneous phase difference across the junction in the experiment, caused by the flux threaded through the junction area and a spatially varying random phase due to the combination of the parallel field and the interface roughness of the superconducting leads[3].

## 7  Majorana Wavefunction

To elucidate possible Majorana signatures in the conductance, we calculate the local density of states (LDOS) associated with the lowest energy state in the junction. A bona-fide Majorana state only exists in a semi-infinite system. To consider this situation, we set $L \to \infty$ in the setup in Figure S14, such that both superconducting leads and the normal region are semi-infinite. We can then calculate the LDOS at zero energy as a function of position using KWANT.

The resulting spectra weight of the Majorana state is shown in Figure S15(a). While the state is indeed localized at the edge, its decay is nonexponential. This is consistent with the fact that the coherence length is very long for the parameters assumed in the previous section. The typical scale for the induced gap is $1/mW^2 \simeq 6\,\mu$eV which agrees with our numerical calculations of the energy spectrum. This gap translates into a coherence length of $\sim 45\mu m$.

We can also calculate the LDOS in finite-length junctions using the setup in Figure S14. In this case, the Majorana states at the two ends hybridize and form a finite energy excitation. In

Figure S15(b) and (c), we show the LDOS corresponding to the lowest energy state for $L = 6\,\mu$m and $L = 900$ nm. In Figure S15(b), the edge localization remains visible. In contrast, the 900 nm junction in Figure S15(c) is too short to allow for any distinction between edge and bulk and the lowest-energy state is homogeneously distributed along the entire length of the junction.

These results suggest that, even though the system is in the topological phase, signatures of Majorana states cannot be distinguished from bulk signatures in devices with current dimensions.

## 8 Origin of the Zero-Bias Conductance Peak

The normal probe in the experiment is in the tunneling regime and the conductance is therefore expected to reflect the local density of states in the junctions. The long Majorana localization length discussed above implies a weak contribution of Majoranas to the conductance, which is washed out due to the energy broadening. Instead, the subgap conductance will be dominated by the band of Andreev states that extend along the entire length of the junction.

In this section, we derive an analytical formula for the Andreev spectrum in a certain limiting case, which can be compared with the numerical results for the full Hamiltonian (1) shown in Figure 1e of the main text. We then use this estimate to calculate the density of states.

To gain analytical insight, we assume that edge effects are small and simply consider the bulk density of states in an infinitely long junction, in which the momentum component $k_x$ is conserved. We first focus on the case of perfect Andreev reflection and later include normal reflection. As

previously, we determine the bound state energies $E$ as a function of $k_x$ from the scattering matrix equation (6) within the continuum model Eq. (1). It is expedient to introduce the polar angles $\theta_i$ defined by $\cos\theta_i = k_x/k_{F,i}$, where $i = 1,2$ denotes the two Fermi surfaces and $k_{F,1/2} = \sqrt{2m\mu} \pm m\alpha$.

We can further simplify Eq. (6) in the case that Andreev reflection does not mix the two Fermi surfaces. This is strictly true only at $k_x = 0$, where the spins on the two Fermi surfaces are antiparallel, while at nonzero $k_x$, a right moving electron can be Andreev reflected into a superposition of left moving holes from both Fermi surface sheets due to a nonzero overlap of spin states[4].

In the limit $E_Z, \Delta \ll \alpha k_F$, however, the two spins along a particular momentum direction are antiparallel because of the Rashba-induced spin-momentum locking. In this case, the two Fermi surfaces at a fixed $k_x$ have approximately opposite spins as long as the corresponding polar angles have similar magnitude $\theta_1(k_x) \simeq \theta_2(k_x) \equiv \theta(k_x)$.

Under these conditions, Eq. (6) yields[4]

$$\arccos\left[\frac{E(k_x)}{\Delta}\right] = \frac{\pi}{2}\frac{E(k_x)}{E_T(k_x)} - \frac{\pi}{2}\frac{\sin\theta E_Z}{E_T(k_x)} \pm \frac{\phi}{2} + n\pi, \qquad n \in \mathbf{Z}, \qquad (10)$$

where $E_T(k_x) = (\pi/2)\sin\theta v_F/W$ is the Thouless energy. The angular dependence of $E_T(k_x)$ reflects the longer distance between the superconducting leads for wavepackets with oblique incidence. The factor $\sin\theta E_Z$ accounts for the fact that spins rotate along the Fermi surface. Considering the limit $E_T(k_x) \gg \Delta$ as is appropriate in our parameter regime for states near $k_x = 0$

14

$[E_T(k_x = 0) = 630 \,\mu\text{eV}]$, we find

$$E = \Delta \cos\left(\frac{WE_Z}{v_F} \pm \frac{\phi}{2} + n\pi\right) \quad (11)$$

which is independent of $k_x$. Hence the bands are expected to be relatively flat. As $k_x$ approaches the Fermi momentum, $E_T(k_x)$ decreases and the energy becomes momentum dependent. This is behavior is consistent with the spectrum in Figure 1e of the main text. We caution, however, that Eq. (10) requires $\theta_1 \simeq \theta_2$ and therefore ceases to be valid as $k_x$ approaches the Fermi momenta.

The flatness of the dispersion should result in a strong peak in the LDOS at the corresponding energy, while the spectral weight at other energies is smaller. This implies that topological phase transitions, where the energy in Eq. (11) vanishes, are signaled by a conductance peak at zero bias. Importantly this is a feature of the bulk density of states and does not take into account the contribution of the Majorana states.

The spectrum changes when normal reflection is taken into account. We specifically consider normal reflection from the superconducting leads that conserves the momentum along the $x$ direction, e.g., caused by a potential step at the N-S interface. In this case, we have to modify Eq. (10) by replacing

$$\phi \to \tilde{\phi} = \arccos[r^2 \cos(2 \sin\theta k_F W + 2\varphi_N) + (1 - r^2)\cos\phi], \quad (12)$$

where $r$ is the normal reflection amplitude and $\varphi_N$ is a phase shift associated with the normal reflection that depends on microscopic details. Details of the derivation can be found in Reference 4 (the prefactor $\sin\theta$ was missing in the reference). Since for our choice of parameters we have

15

$k_F W \gg 1$, the first term in the square brackets above oscillates rapidly as a function of $k_x$. This results in oscillation of the Andreev energy as a function of $k_x$ with an amplitude that scales with the reflection probability.

Indeed, these oscillations also occur in the numerical spectra shown in Figure 1f of the main text.

Since normal reflection broadens the band of Andreev states, the density of states calculated from Eqs. (11) and (12) exhibits significant contributions over a range of energy values. This can be seen in Figure S16, which shows the density of states for various values of the in-plane magnetic field calculated using Eqs. (11) and (12).

There are two van Hove singularities at the band edges because all maxima and minima of $E(k_x)$ in Eq. (11) occur at the same energies.

The band center and the approximate band width in Figure S16a is consistent with the numerical spectra shown in Figure 1e of the main text. Moreover, for $B = 0\,\text{T}$ and $1.6\,\text{T}$ the energy minima and maxima of the spectra are roughly aligned for not too large momenta in agreement with our analytical findings. The spectrum at $B = 1\,\text{T}$, in contrast, has aligned minima but not maxima.

By varying either $\phi$ or $E_Z$, one can tune the band center across zero energy. Once the band crosses zero energy the positive and negative energy band overlap which leads to an enhanced

16

density of states at zero energy. For weak normal reflection, the band widths are small and the density of states shown in Figure S16a is enhanced near zero energy only in a narrow window around the phase transition. This behavior is consistent with the spectra shown in Figure 1e of the main text.

If the normal reflection is higher, the band width increases as shown in Figure S16b. In this case, the density of states is enhanced near zero energy in a larger range of magnetic fields. While the density of states does not typically have a maximum at zero, an additional energy broadening applied to the data may result in a zero-bias peak. If the broadening is sufficiently large, the peak can persist between $B = 0.8\,\text{T}$ and $B = 1.2\,\text{T}$ for the parameters of Figure S16b. Such a behavior is quantitatively consistent with the numerical data for the conductance in Figure 4g of the main text, which shows a peak at zero bias over an extended region in parameter space.

We conclude that a zero-bias peak in the conductance signals the zero-energy crossing of a band of Andreev states, which corresponds to a topological phase transition. The width of the band is determined by the normal reflection probability from the superconducting leads. In the presence of a nonzero band width and an additional broadening of the data in energy, a zero-bias peak can occur over an extended range of magnetic field strengths and phase differences in agreement with the experimental observations as well as the numerical results presented in the main text.

## 9 Finite-Difference Scheme and Exact Diagonalization

To show that the peak in the $dI/dV$ characteristics at the topological transition is a universal property for wide junctions, while for narrower junctions a competition between Andreev and Majorana bound states occurs, we perform a second type of calculations. Complementary to the previous section, we employ a finite-difference scheme to obtain the eigenspectrum and eigenstates of the finite Josephson junction shown in Figure S17. These eigenstates and eigenenergies are then used to compute the local density of states (LDOS) at the edge of the normal region. For this purpose, we use the Hamiltonian (Eq. 1-4) and invoke hard-wall boundary conditions in $x$- and $y$-directions to describe a Josephson junction of finite length $L$ in $x$-direction and finite width $W_{tot} = W + 2W_{SC}$ in $y$-direction. Here, $W$ is the width of the normal region and $W_{SC}$ denotes the finite widths of the superconducting regions (see Figure S17). We discretize the Hamiltonian (Eq. 1-4) using $d\psi(x)/dx \to [\psi(x_i+a)-\psi(x_i-a)]/2a$, $d^2\psi(x)/dx^2 \to [\psi(x_i+a)-2\psi(x_i)+\psi(x_i-a)]/a^2$, and analogous expressions for the derivatives with respect to $y$, where $a$ and $x_i = ai$ (with $i \in \mathbb{Z}$) denote the discrete step size and lattice sites, respectively. This procedure yields Eq. 5. If Eq. 5 is written in the form

$$H = \sum_{i,i'=1}^{L} \sum_{j,j'=-W_{SC}+1}^{W+W_{SC}} \mathcal{H}_{i,j;i',j'} \, c_{i,j}^\dagger c_{i',j'}, \tag{13}$$

the eigenenergies $E^{(n)}$ and eigenstates $c_{i,j}^{(n)}$ of the system are found by diagonalizing the eigenvalue equation

$$\sum_{i'=1}^{L} \sum_{j'=-W_{SC}+1}^{W+W_{SC}} \mathcal{H}_{i,j;i',j'} \, c_{i',j'}^{(n)} = E^{(n)} c_{i,j}^{(n)}, \tag{14}$$

where $n$ denotes the $n$-th eigenstate.

These eigenenergies and eigenstates are then used to compute the LDOS at the upper edge of the normal region as a function of energy $E$,

$$\text{LDOS}(E) = \sum_{i=1}^{l_x} \sum_{j=W/2-l_y/2}^{W/2+l_y/2} \sum_{n} \left| c_{i,j}^{(n)} \right| \delta_\Gamma(E - E^{(n)}). \tag{15}$$

Here, $\delta_\Gamma(E - E^{(n)})$ is modeled by a Gaussian with broadening $\Gamma$. The area of summation is chosen to resemble the area covered by the tunneling probe with lengths $l_x$ and $l_y$ in $x$- and $y$-directions, respectively (see Figure S17). In the lowest-order approximation, the $\mathrm{d}I/\mathrm{d}V$ characteristic is proportional to the LDOS. Hence, to compare with the experiment and the KWANT simulations, we first compute the LDOS for a given probe area and position and then look at the curvature $\partial^2 \text{LDOS}/\partial E^2$ around $E = 0$.

## 10 Comparison of the Local Density of States for Josephson Junctions with Different Widths

In general, a zero-energy peak in the edge LDOS can originate from Majorana end states and/or low-energy bulk Andreev bound states. As discussed in the main text, the most pronounced zero-energy peaks for a 600nm wide junction are expected to actually arise from the bulk Andreev bound states that describe the topological phase boundaries. Here, we study and compare the coexistence of zero-energy peaks in the edge LDOS due to Majorana end states and bulk Andreev states for samples of width $W = 600$ nm and $W = 400$ nm.

For the calculations with a $W = 600$ nm wide normal region, we have used $\alpha$=16 meVnm,

$\Delta$=65 $\mu$eV, $\mu_N = 12.5$ meV, $\mu_{SC} = 25$ meV, $g_N = 10$, $g_{SC} = 0$, and $m = 0.038 m_e$. The size of the superconducting regions is 900nm×700nm and the size of the normal region 900nm×600 nm in $x$- and $y$-directions, with a probe size of 100nm×100nm. The discretization step is chosen as $a = 10$ nm. The lengths and widths as well as their corresponding numbers of steps are described by the same quantities, i.e. $L = 90$, $W_{tot} = 200$, $W_{SC} = 70$, $W = 60$, $l_x = 10$, and $l_y = 10$. To compute the LDOS via Eq. (15), a Gaussian broadening of $\Gamma = 0.25\Delta$ for $\delta_\Gamma(E - E^{(n)})$ has been used.

Figure S18 shows $\partial^2 \text{LDOS}/\partial E^2$ at $E = 0$ for the above parameters as a function of the magnetic field $B_x$ and the external phase difference $\phi$ between the two superconductors. Similar to the experiment and our KWANT simulations, we find that a zero-energy peak ($\partial^2 \text{LDOS}/\partial E^2 < 0$) in the LDOS starts to appear at fields around $B_x = 0.4 - 0.5$ T and becomes the most pronounced around $B_x = 1.2$ T. As $B_x$ increases further and $(\phi, B_x)$ moves away from the boundaries of the topological phase, the zero-bias peak disappears.

We note that we have chosen somewhat different parameters for the superconducting regions here compared to the KWANT simulations: The finite width $W_{SC}$ of the superconducting regions is reduced to 700 nm (compared to 1200 nm), thereby increasing the amount of normal reflection due to the finite size of the superconductors. This effect is somewhat compensated for by reducing the mismatch between the chemical potentials $\mu_N$ and $\mu_{SC}$ (12.5:25 compared to 12.5:50 in the KWANT calculations), which is a second source for normal reflection at the S/N interfaces.

Likewise, the strength of Rashba SOC in the entire device is reduced by a factor of around 2

to $\alpha = 16$ meVnm to demonstrate the robust occurrence of a zero-energy peak in the LDOS/a zero-bias peak in the $\mathrm{d}I/\mathrm{d}V$-characteristic for a wide range of Rashba SOC strengths. This robustness with Rashba spin-orbit strength $\alpha$ of the topological phase boundaries and of zero-energy peaks at (segments of) these boundaries can also be explained within a scattering matrix approach.

The nature of the Majorana bound state is illustrated in Figure S19, which shows the probability amplitude $|\Psi(x,y)|^2$ of the lowest-energy state at $B_x = 1.2$ T and a phase difference $\phi = \pi$. Due to the finite length $L$ of the setup and the resulting hybridization of the two separate Majorana end states, this state is not situated at zero-energy. As argued above in Supplementary Section 7, due to the small induced topological gap of the order of 6 $\mu$eV protecting the Majorana end states, these states are expected to decay only slowly away from the upper and lower edges of the normal region. This feature is exhibited in Figure S19, where the lowest energy state is spread over the entire normal region and only slightly localized at the boundaries of the normal region (see also Figure S15).

The large localization length of the Majorana end states in the sample with a 600-nm-wide normal region implies that the zero-energy peaks in the edge LDOS arise almost entirely from the bulk Andreev bound states at the topological phase boundaries. Indeed, comparing $\partial^2\mathrm{LDOS}/\partial E^2$ around $E = 0$ computed at the edge with that computed in the center of the normal region yields qualitatively similar results.

To obtain more localized Majorana bound states, the topological gap, of the order of $\hbar^2/(mW^2)$, should be increased. Hence, a potential route is decreasing the width $W$ and going to narrower de-

21

vices. In fact, a sample with a $W = 400$ nm wide normal region is also available, Device Kappa in Supplementary Section 5. Figure S20 shows $\partial^2\text{LDOS}/\partial E^2$ obtained for a junction of width $W = 400$ nm. In Figure S20, the parameters are $\alpha$=160 meVnm, $\Delta$=130 $\mu$eV, $\mu_N = 10$ meV, $\mu_{SC} = 20$ meV, $g_N = 10$, $g_{SC} = 0$, and $m = 0.038 m_e$. Here, the size of the superconducting regions is 1000nm×800nm in $x$- and $y$-directions, that of the normal region 1000nm×400 nm, and that of the probe 100nm×100nm. Using again a discretization step of $a = 10$ nm, the corresponding numbers of steps are $L = 100$, $W_{tot} = 200$, $W_{SC} = 80$, $W = 40$, $l_x = 10$, and $l_y = 10$ (again these symbols denote widths/lenghts as well as the number of steps). The LDOS is computed with a Gaussian broadening of $\Gamma = 0.2\Delta$ in Eq. (15).

Similarly to Figure S18, the curvature of the LDOS at the edge of the normal region is shown in Figure S20(a). For these parameters, the topological phase arises only for fields above $B_x = 1$ T, as illustrated by the dashed lines denoting the boundaries of the topological phase. These boundaries have been obtained numerically for an infinitely long system ($L \to \infty$), which has otherwise the same parameters as Figure S20(a). Whereas for the edge LDOS, no zero-energy peak is found outside the topological region ($\partial^2\text{LDOS}/\partial E^2 > 0$; red areas in Figure S20), a zero-energy peak of the edge LDOS appears for a wide range of phase differences $\phi$ inside the topological region ($\partial^2\text{LDOS}/\partial E^2 < 0$; blue areas in Figure S20). This is in contrast to the bulk LDOS measured in the center of the normal region (yellow area in Figure S17): As shown in Figure S20(b), no zero-energy peak develops for the bulk LDOS, apart from a small region in phase space slightly below $B_x = 3$ T and from $\phi \approx 0$ to $\phi \approx 0.4\pi$.

These regions with $\partial^2 \text{LDOS}/\partial E^2 < 0$ in the bulk LDOS arise from Andreev bound states at the phase boundary, providing again a fingerprint of the topological transition. We note that due to the much stronger Rashba SOC chosen here compared to Figure S18, the radii of the two Rashba-split Fermi circles differ considerably. Thus, the Andreev bound states arising mainly from the larger of the two Fermi circles carry significantly more weight in the LDOS than those arising from the smaller Fermi circle. As a consequence, pronounced zero-energy peaks in the bulk LDOS occur primarily at those phase boundaries that correspond to Andreev bound states formed from the larger Fermi circle.

Moreover, Figure S20(b) shows that the phase boundaries are usually traced by regions of nearly zero curvature (white regions), indicative of a flat LDOS at these boundaries. This is consistent with the high normal reflection that is implied by the squeezed topological regime in Figure S20. Indeed, large normal reflection increases the band widths of the Andreev bound states and thus the range of energies over which a flat LDOS can be expected (see Sec. 8). For the energy broadening $\Gamma = 0.2\Delta$ in Figure S20, the LDOS around $E = 0$ remains flat at the phase boundaries.

The above difference between the edge and bulk LDOS points to the zero-/low-energy states being localized at the edge of the normal region. This is indeed corroborated by Figure S21, which, as an example, shows the wave function of the lowest-energy state at $B_x = 2.4$ T and $\phi = 1.8\pi$: Here, the weight of the low-energy wavefunction at the edge of the normal region is significantly higher than the weight in the center of the junction.

The above results imply that for junctions with normal regions of width $W = 400$ nm, it

23

could be possible to observe signatures of Majorana bound states by comparing the edge and bulk LDOS. Such signatures, however, are expected to occur for magnetic fields beyond the current experimental regime. Still, going to narrower devices with improved fabrication technology offers a promising route for finding fingerprints of Majorana bound states, difficult to isolate in Josephson junctions with wide normal regions.

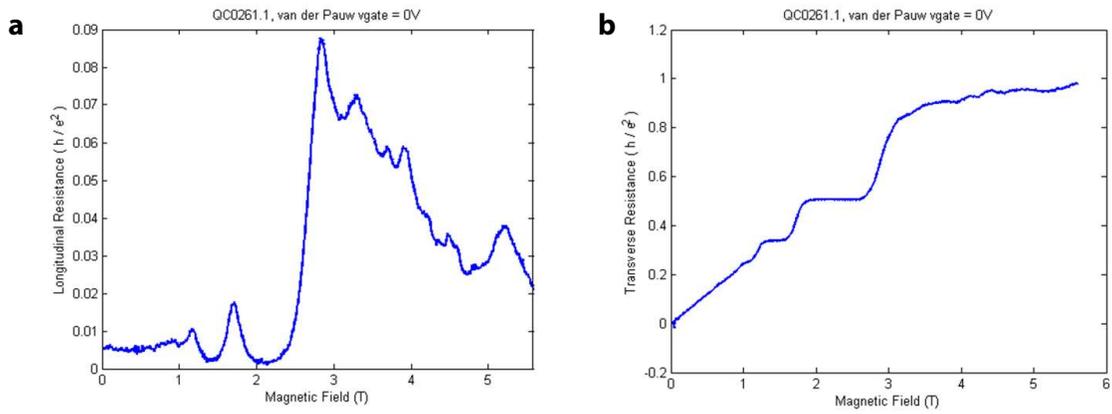

**Figure S1  Quantum Hall effect in the HgTe quantum well QC0261.** Shown here are the longitudinal and Hall resistances of a van der Pauw device fabricated on wafer QC0261, as a function of the perpendicular magnetic field, at a top gate voltage of 0 V, measured at a temperature of 4 Kelvin.

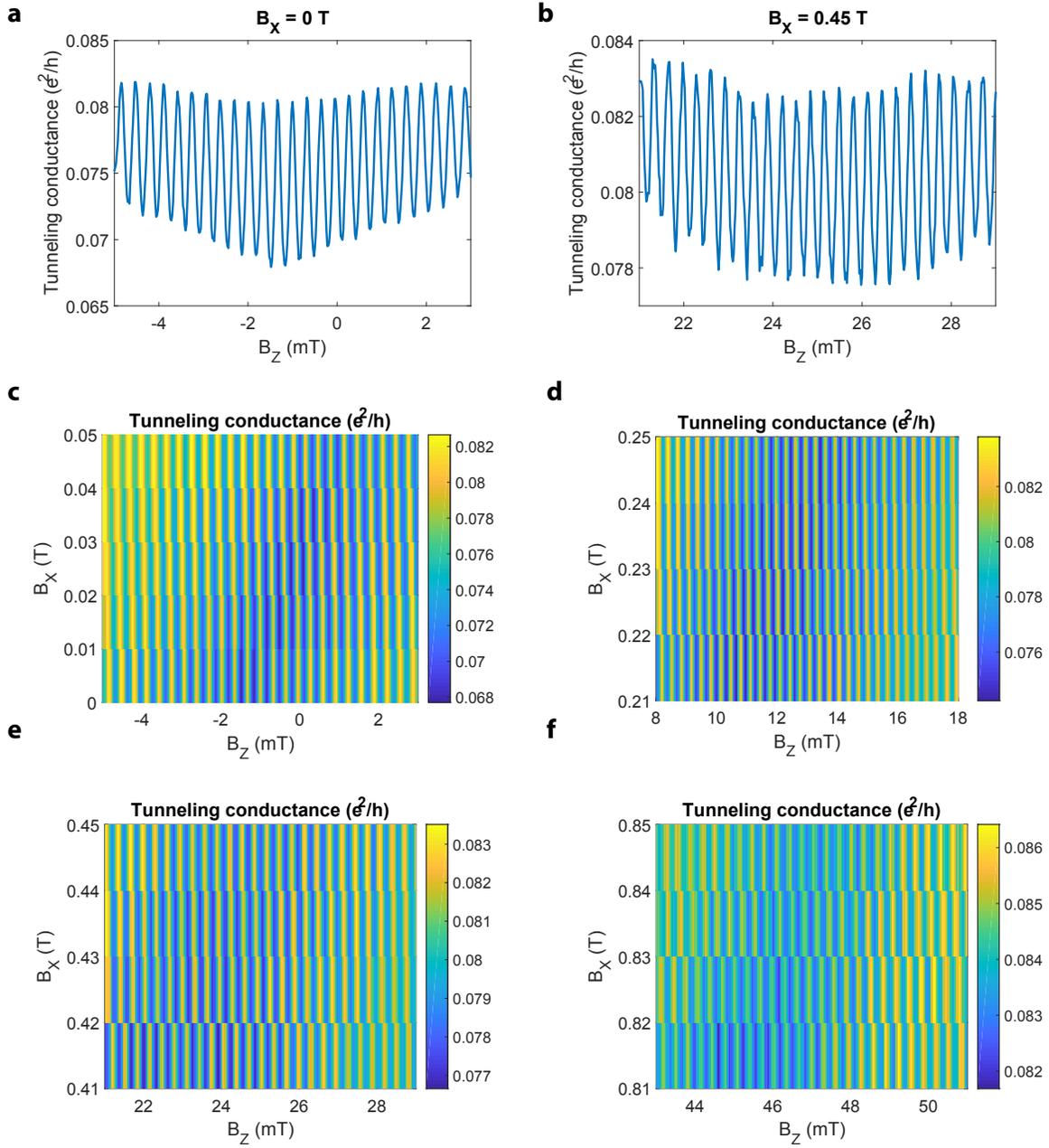

**Figure S2 Compensation for the sample-magnet misalignment.** a and b, The tunneling conductance near zero bias as a function of the out-of-plane magnetic field, $B_z$, at zero in-plane field $B_x = 0$ T in a) and a finite in-plane field $B_x = 0.45$ T in b). c-f, Colourplots of tunneling conductance near zero bias as a function of both the out-of-plane magnetic field $B_z$ and the in-plane field $B_x$. At incremental ranges of $B_z$, these uncompensated plots show the linear offset of $B_z$ growing with increasing $B_x$.

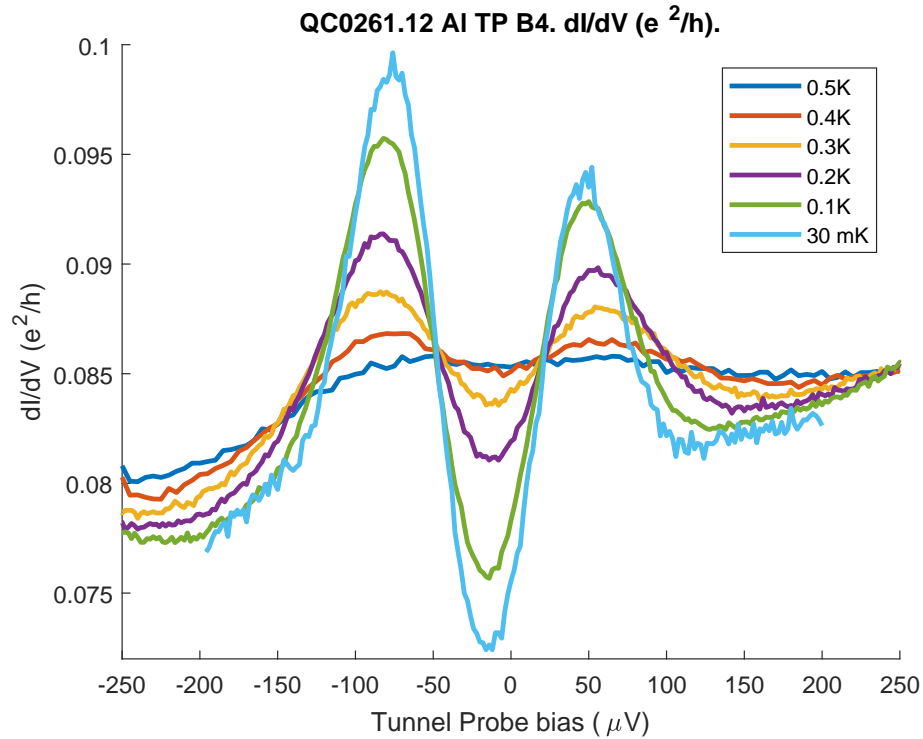

**Figure S3  Temperature dependence of the tunneling conductance.** Shown here is temperature dependence of the tunneling conductance of the same device as discussed in the main text, plotted against the DC bias across the tunnel probe and the superconducting lead. Different curves represent data taken at different averaged temperatures during its cool-down in our dilution refrigerator, from 0.5 Kelvin (blue curve) down to 31 millikelvins (teal curve) over a range of 0.5 mV of DC bias.

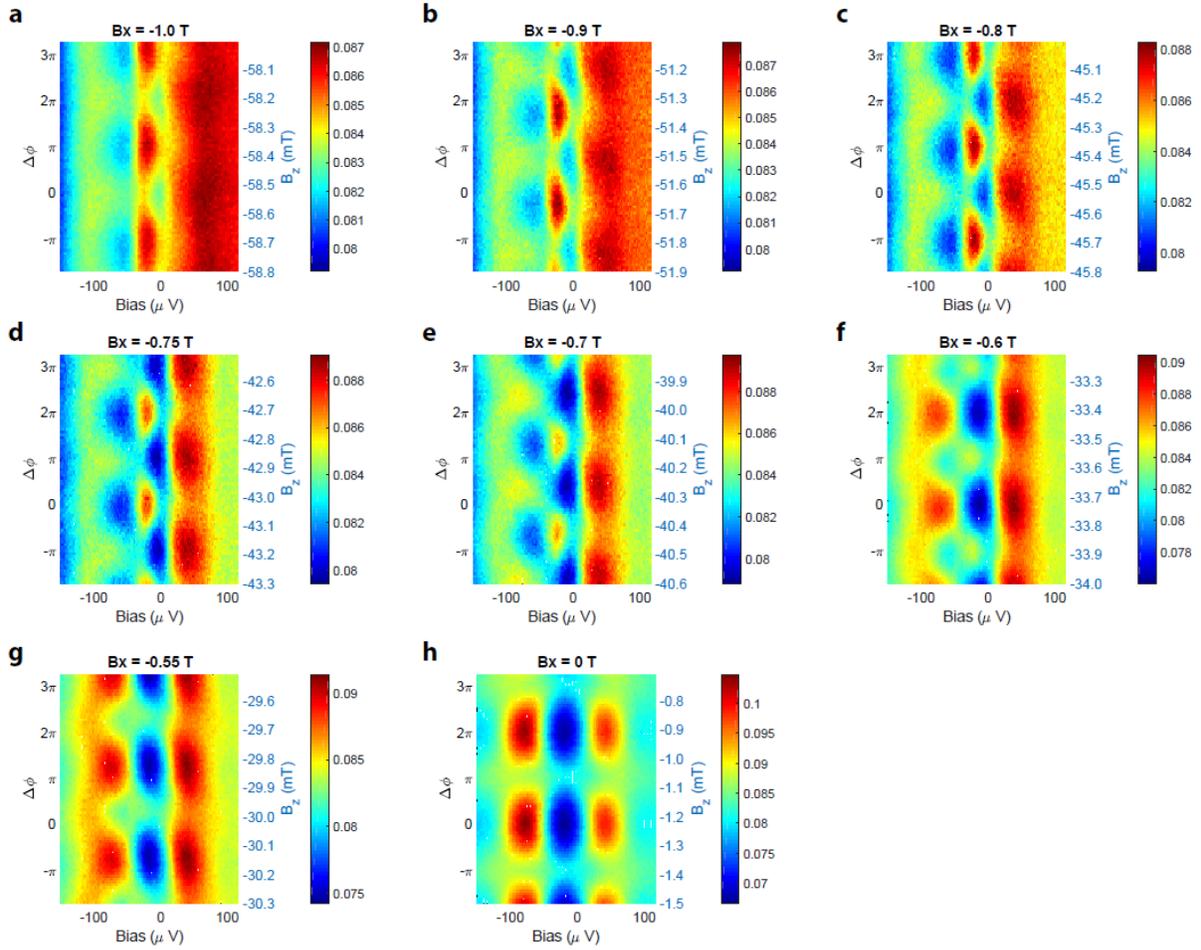

**Figure S4 Unprocessed conductance data with in-plane magnetic fields, part 1.** The raw conductance data (in units of $e^2/h$) at in-plane fields ranging from -1.0 T to 0 T, with an electron-hole asymmetry which grows as the magnitude of the in-plane field increase from 0 to 1.0 T. All plots show tunneling conductance in units of $e^2/h$ as a function of both bias voltage and the span over a range of 0.8 mT in $B_z$.

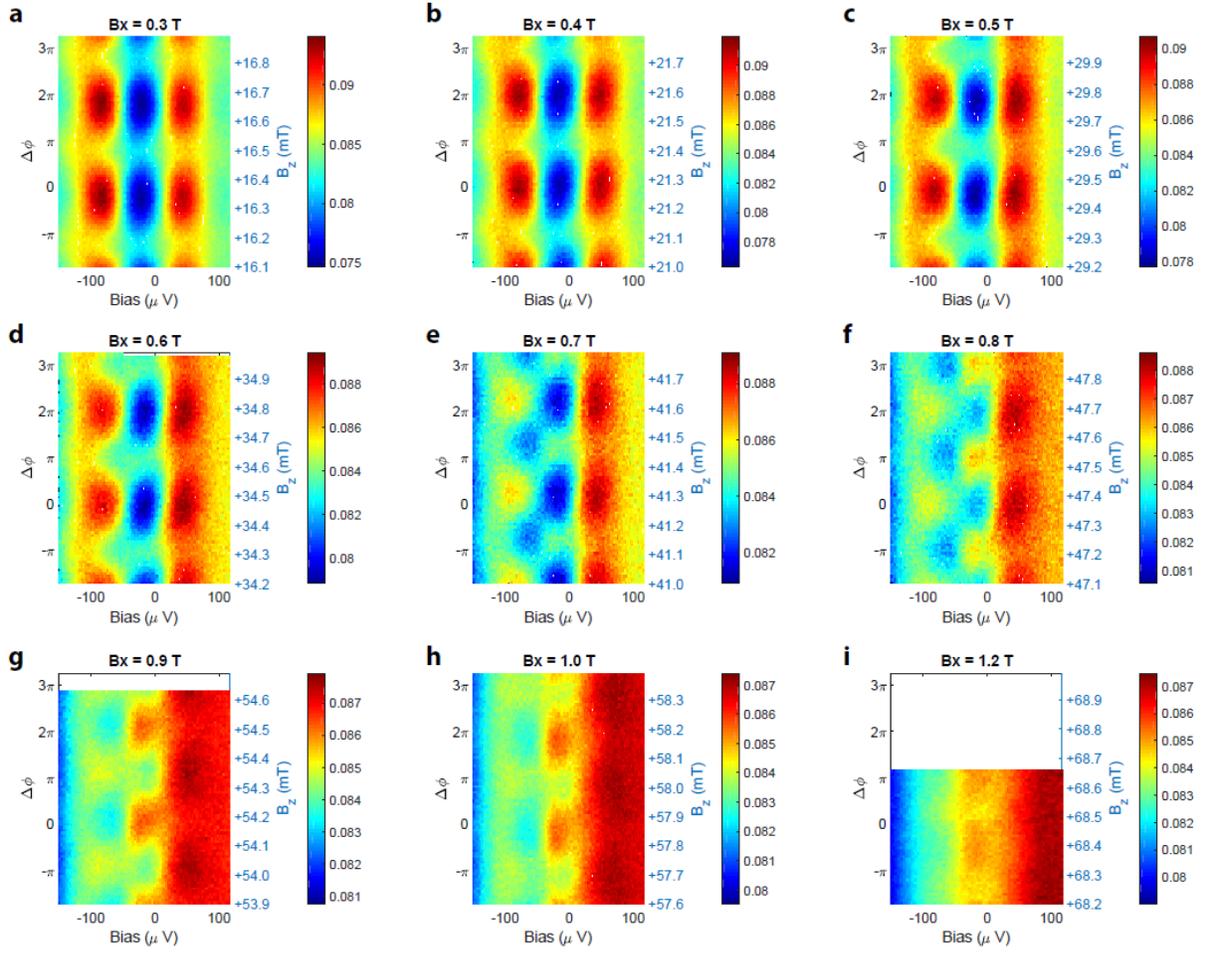

**Figure S5   Unprocessed conductance data with in-plane magnetic fields, part 2.** The raw conductance data (in units of $e^2/h$) with an electron-hole asymmetry which grows as the in-plane field increases from 0.3 T to 1.2 T. All plots show tunneling conductance in units of $e^2/h$ as a function of both bias voltage and the span over a range of 0.8 mT in $B_z$.

30

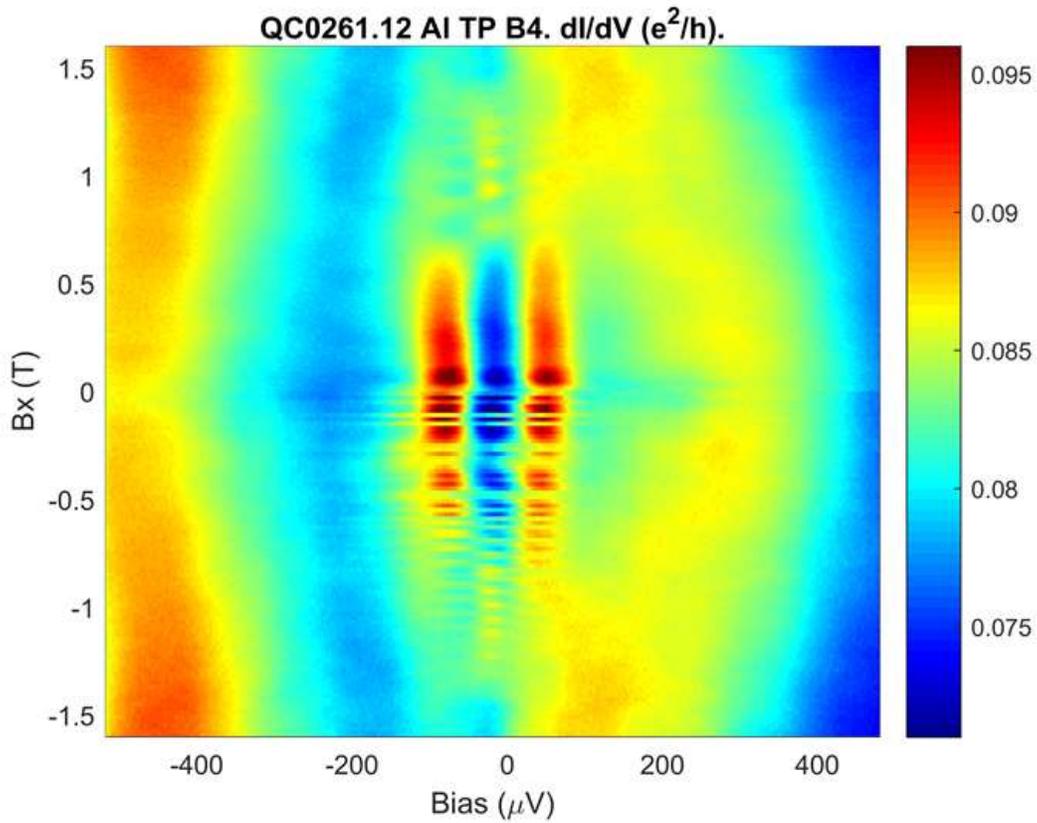

**Figure S6  Differential conductance colourmap of device Beta as $B_x$ varies continuously.** Shown here is a continuous evolution of the tunneling conductance curve, over a wide range of the in-plane field $B_x$ from -1.6 T to 1.6 T. Because our compensation of the perpendicular component of the in-plane magnetic field is not perfect, there is still a small flux which causes oscillations in the phase $\phi$ at the same time as the field grows. Data shown here is obtained from the same device as discussed in the main text.

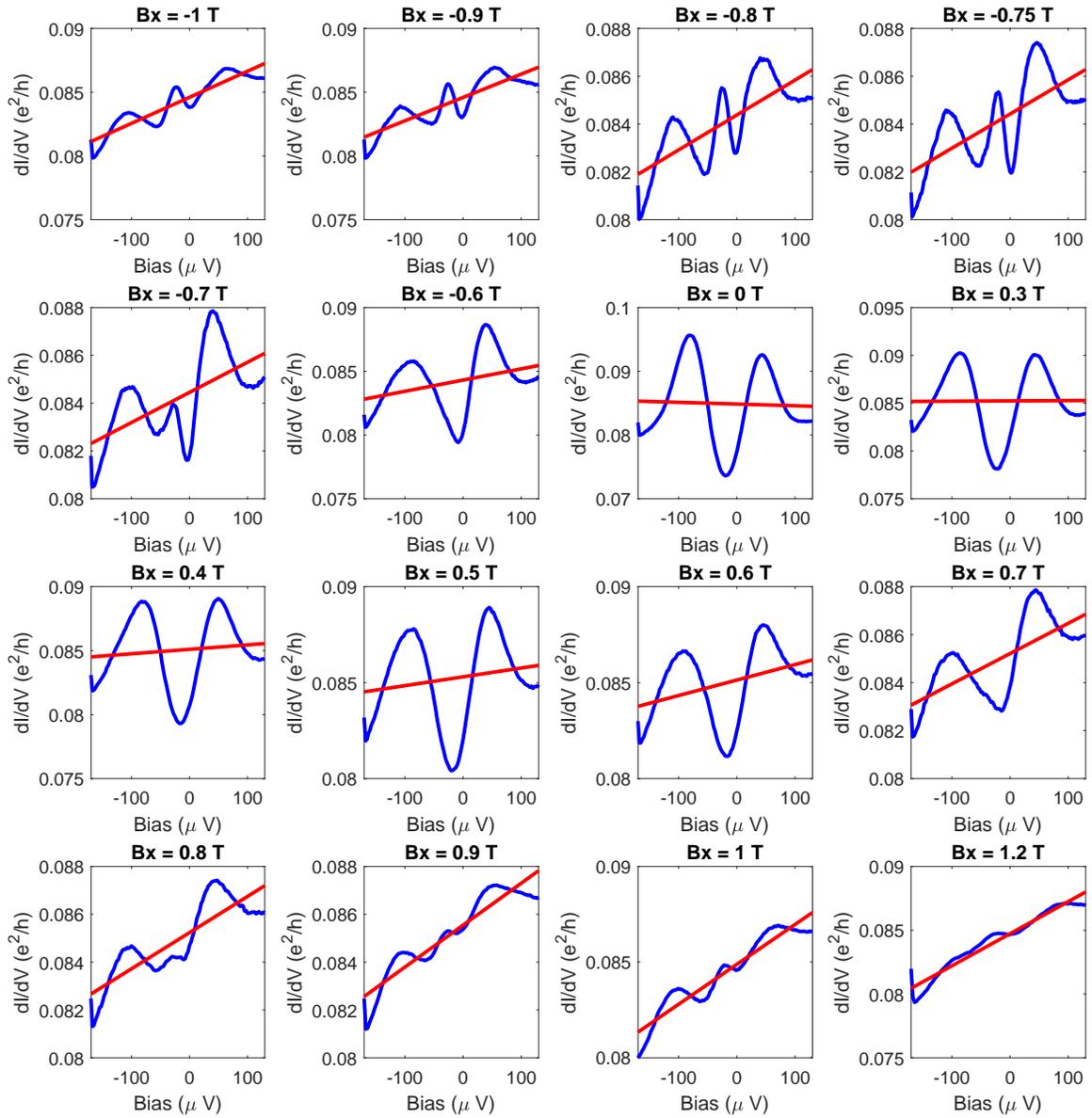

**Figure S7 Development of the electron-hole asymmetry with the in-plane field.** Each panel above shows, in a blue, solid curve, the raw measurement of the tunneling conductance, averaged over an entire period in the phase difference $\phi$, at a given in-plane field $B_x$, labeled above the panel. The red line in each panel shows the linear regression of that tunneling conductance curve, from which the slope representing the asymmetry is extracted.

32

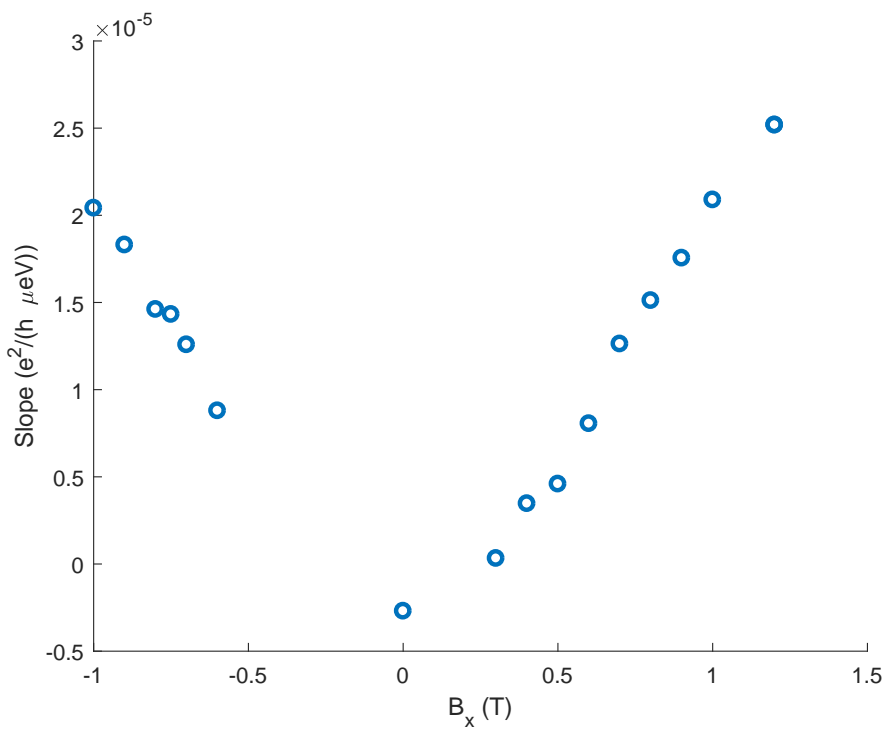

**Figure S8  Slope of the electron-hole asymmetry versus the in-plane magnetic field.** Each point in the above plot represents a slope extracted from the linear regression shown in Figure S7, in units of $e^2/(h \cdot \mu eV)$. Overall, the asymmetric slope increases linearly with the magnitude of the in-plane field, and it is independent of the sign of the field.

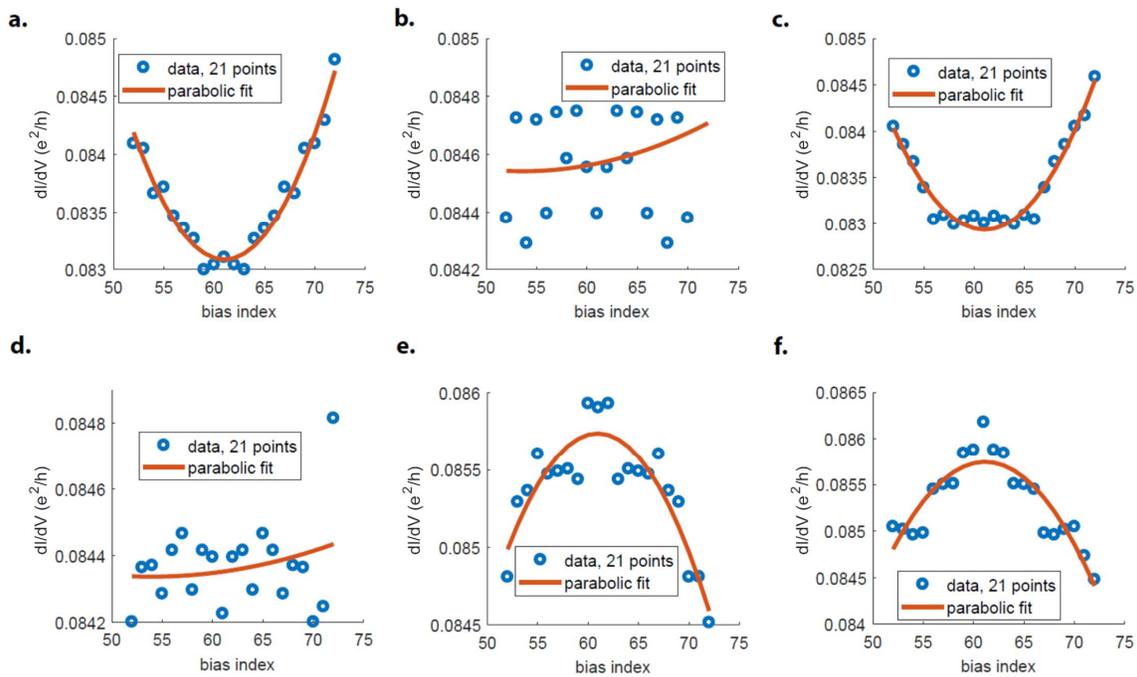

**Figure S9  Extracting the curvature of zero-energy peaks using parabolic fits.** Shown above are a few examples of extracting the curvature of zero-energy peaks using parabolic fits, at $B_x$ =0.8 T, demonstrating the fitting procedure works nicely for a well-defined peak (dip), producing a negative (positive) curvature. Yet, the parabolic fit works not so well when the curve is flat in this region, in which case it yields a small curvature, as desired.

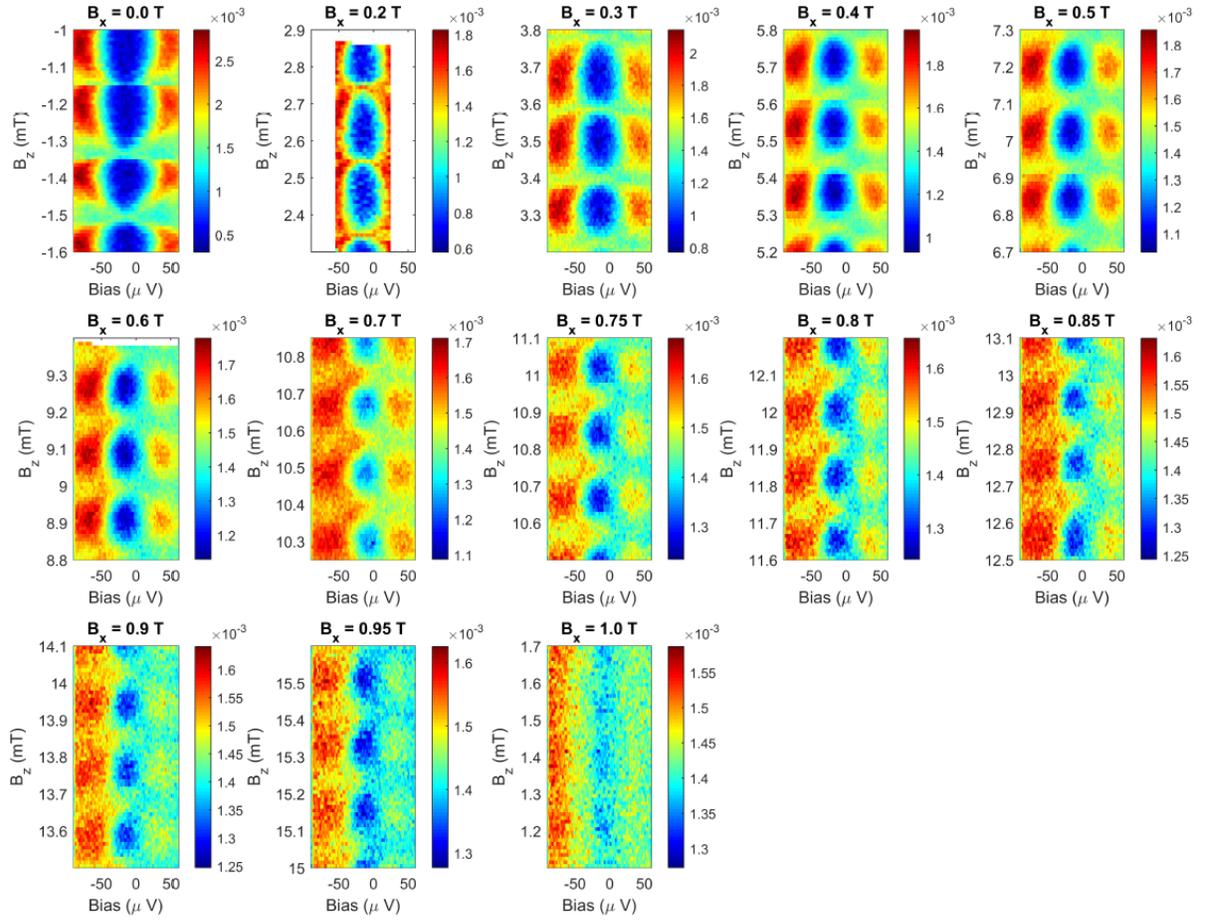

**Figure S10  Differential conductance colourmaps of device Kappa at various $B_x$ values.** Shown here is the raw data for a series of two-dimensional tunneling conductance maps as a function of the bias voltage and the perpendicular field $B_z$, as the in-plane field $B_x$ increases from 0 T to 1.0 T. All plots are in units of $e^2/h$ and span over a range of 0.6 mT in $B_z$.

35

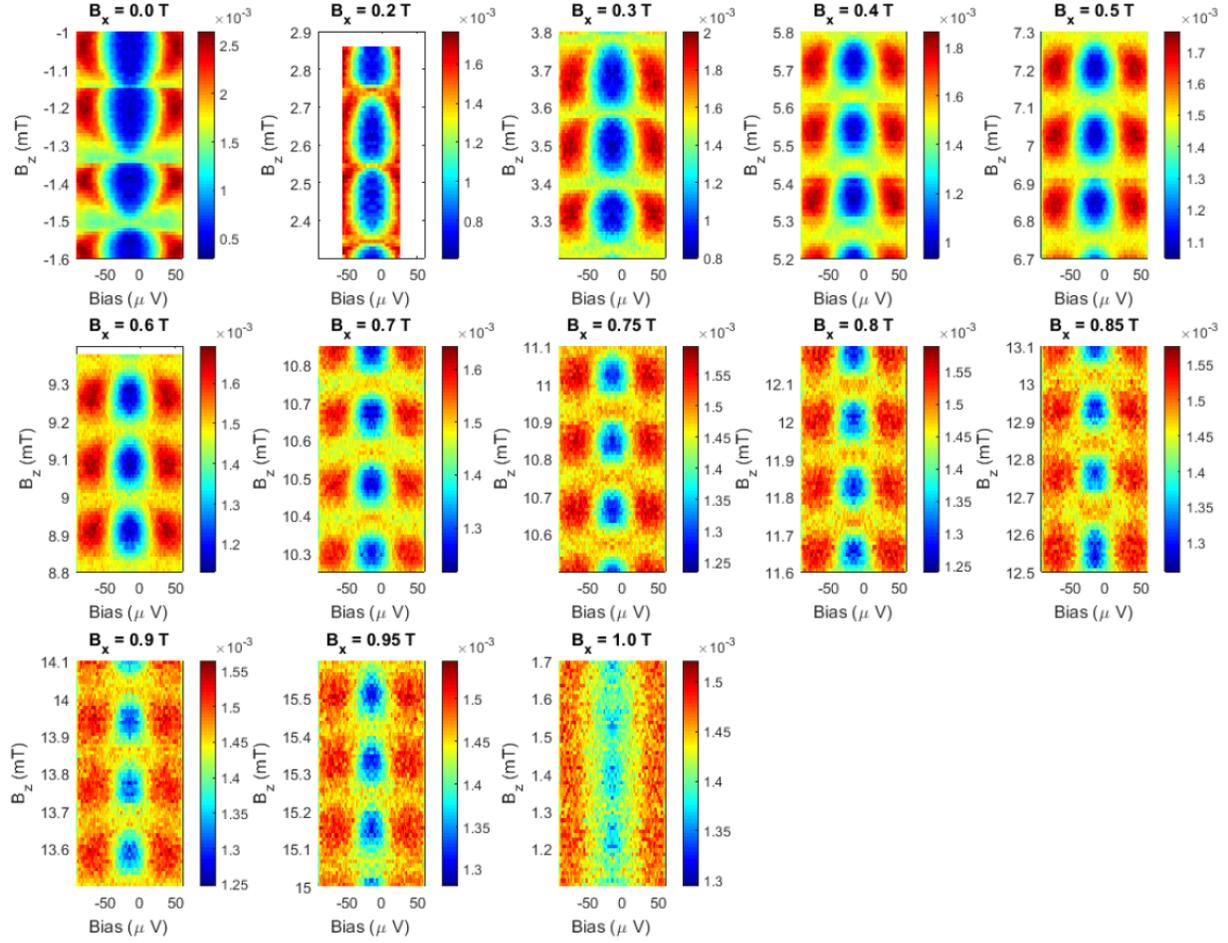

**Figure S11 Symmetrized differential conductance colourmaps of device Kappa at various $B_x$ values.** Shown here is the symmetrized data for a series of two-dimensional tunneling conductance maps as a function of the bias voltage and the perpendicular field $B_z$, as the in-plane field $B_x$ increases from 0 T to 1.0 T. All plots are in units of $e^2/h$ and span over a range of 0.6 mT in $B_z$.

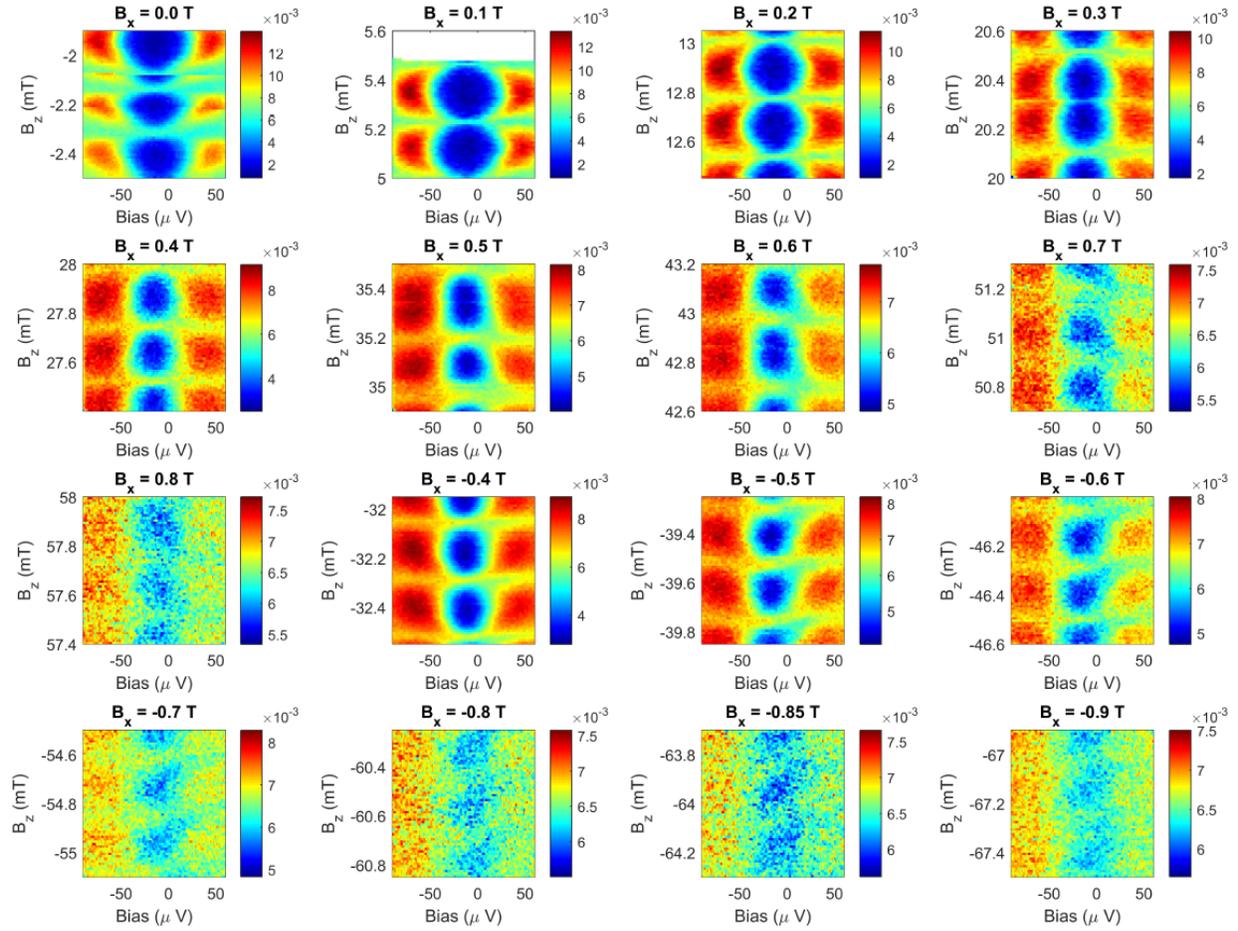

**Figure S12   Differential conductance colourmaps of device Zeta at various $B_x$ values.** Shown here is the raw data for a series of two-dimensional tunneling conductance maps, as the in-plane field $B_x$ ranges from -0.9 T to 0.8T. All plots are in units of $e^2/h$ and span over a range of 0.6 mT.

37

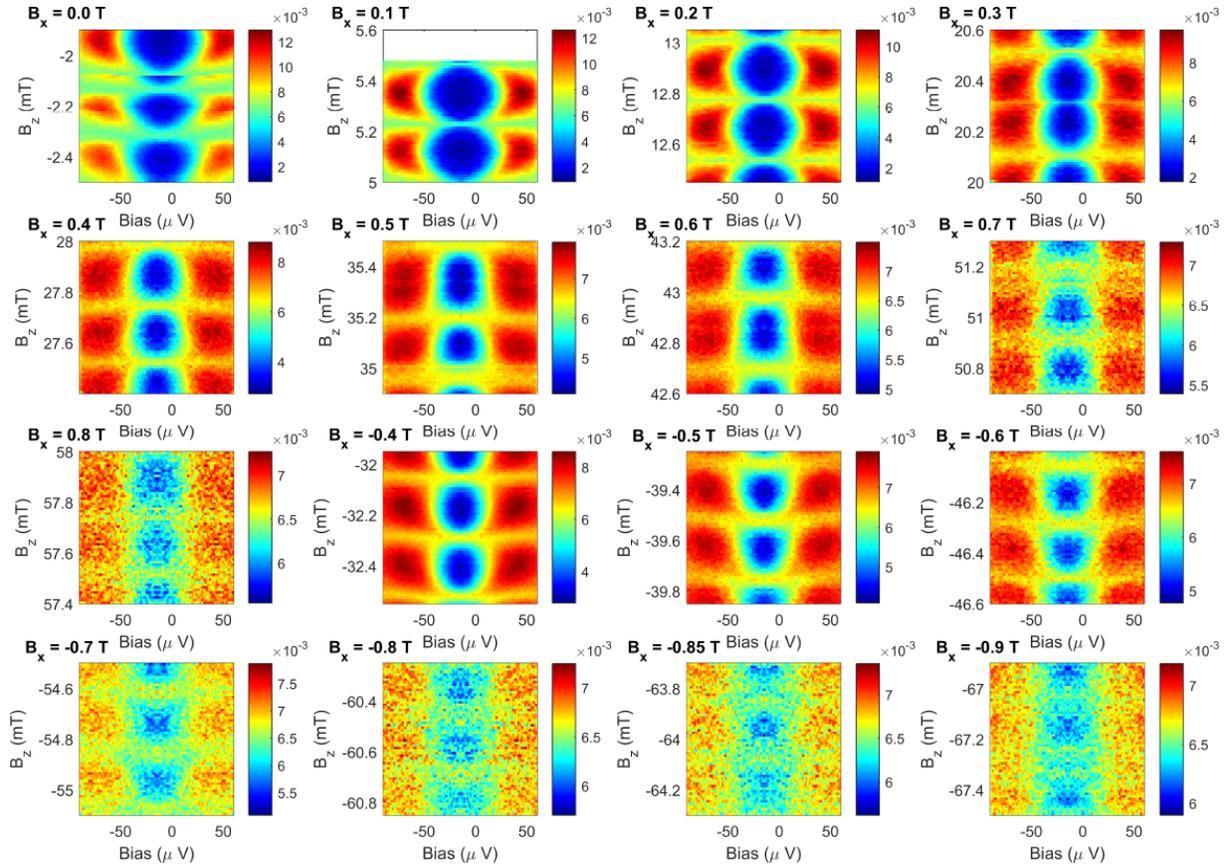

**Figure S13 Symmetrized differential conductance colourmaps of device Zeta at various $B_x$ values.** Shown here is the symmetrized data for a series of two-dimensional tunneling conductance maps, as the in-plane field $B_x$ ranges from -0.9 T to 0.8T. All plots are in units of $e^2/h$ and span over a range of 0.6 mT in $B_z$.

38

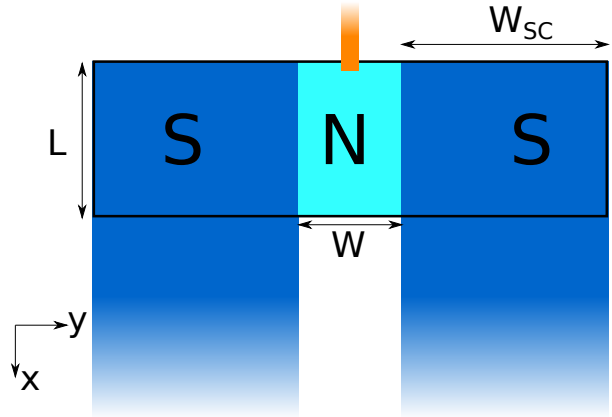

**Figure S14  The setup modeled in the numerical calculations.** The system comprises a rectangular region with a normal and two superconducting segments (black contour). To calculate the conductance, we add a semi-infinite normal lead in the junction center near the upper edge (orange) and two semi-infinite superconducting leads contacting the superconducting segments at the bottom edge.

39

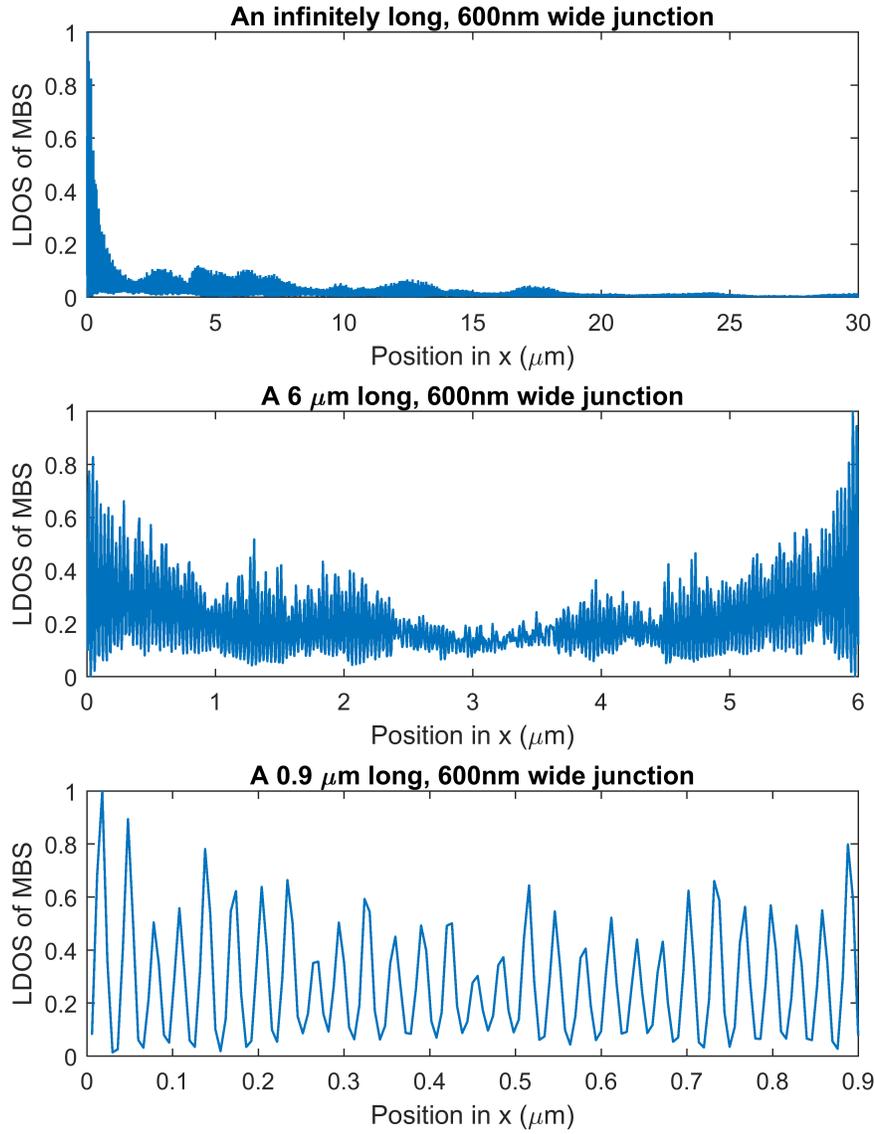

**Figure S15  Delocalization of the Majorana bound state.** Normalized local density of states of the lowest energy excitation, averaged over the width of the junction, as a function of distance from the edge in the setup of Figure S14 with $W = 600\,\text{nm}$ and (a) $L = \infty$, (b) $L = 6\,\mu\text{m}$, and (c) $L = 900\,\text{nm}$.

40

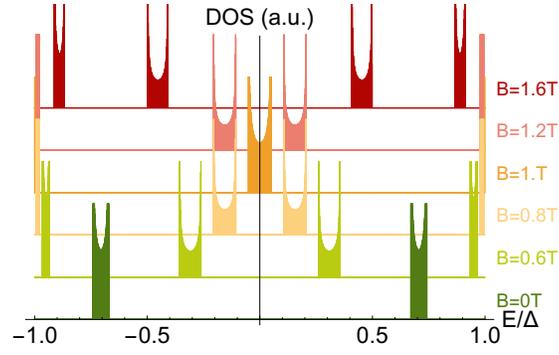

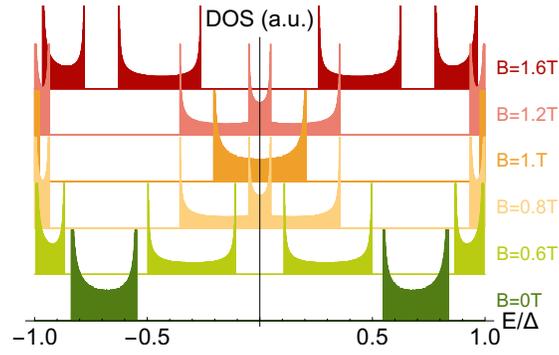

**Figure S16  Calculated density of states of the Andreev band.** Density of states of the Andreev band calculated from the approximate expression in Eqs. (11) and (12) for various values of the magnetic field. The superconducting phase difference is chosen as $\phi = \pi/2$ and the normal reflection probability is a) $r^2 = 0.1$ and b) $r^2 = 0.4$. We assume $k_F W \gg 1$ so that the DOS is independent of $\varphi_N$. The other parameters are chosen as above. This choice implies that the phase transition for $\phi = \pi/2$ and $r = 0$ occurs at $B = 1\,\text{T}$ where $E_Z = E_T/2$.

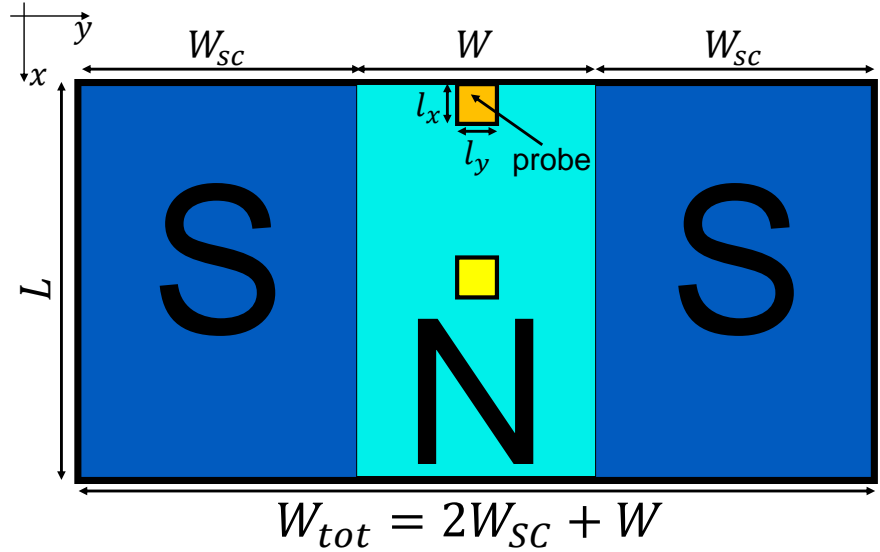

**Figure S17  Setup considered for the complementary finite-difference eigenstate calculations.** The system consists of a normal region of dimensions $900$ nm $\times 600$ nm and two adjacent superconducting regions of dimensions $900$ nm $\times 700$ nm. A phase difference of $\phi$ is applied between the two superconductors. The edge local density of states (LDOS) is calculated for an area of dimensions $100$ nm $\times 100$ nm at the top edge of the normal region (orange area). The LDOS in the bulk is computed from a $100$ nm $\times 100$ nm square in the center of the normal region (yellow area).

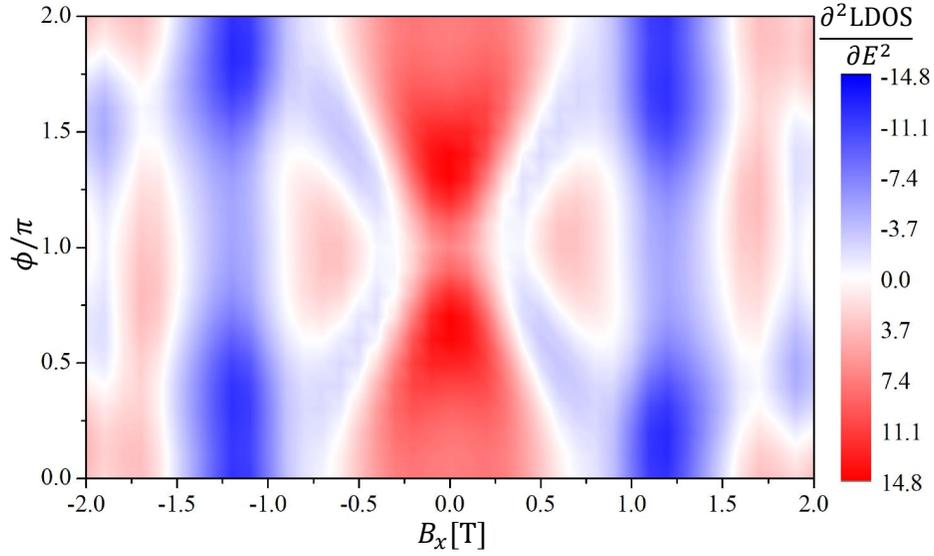

**Figure S18 Predicted zero-bias curvature for a 600-nm junction.** Calculated $\partial^2\mathrm{LDOS}/\partial E^2$ (in a.u.) at $E = 0$ as a function of magnetic field $B_x$ and the phase difference $\phi$ between the superconducting regions for the setup illustrated in Figure S17. The area over which the LDOS is computed corresponds to a probe size of $100$ nm $\times 100$ nm at the upper edge of the normal region (orange area in Figure S17).

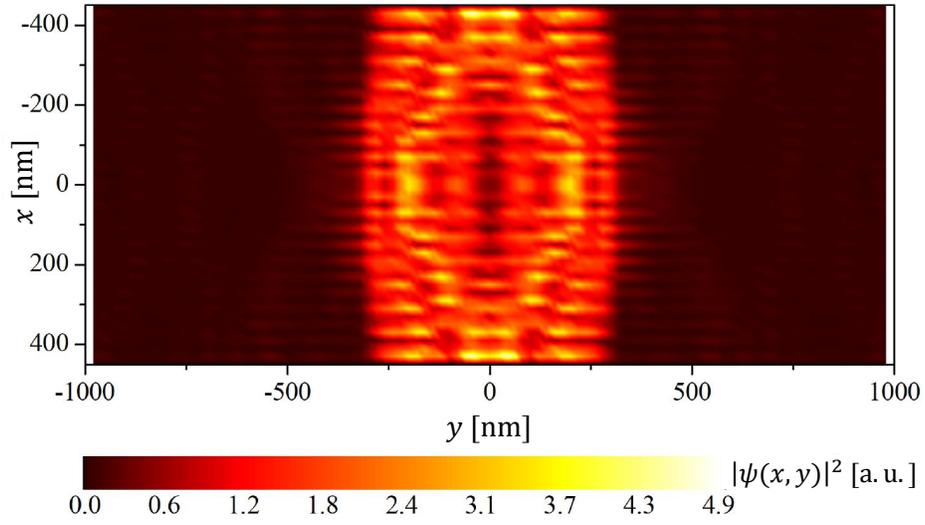

**Figure S19   Wavefunction of the low-energy state for a 600-nm junction.**

Probability amplitudes $|\Psi(x,y)|^2$ of the low-energy state, calculated at $B_x = 1.2$ T and

$$\phi = \pi.$$

44

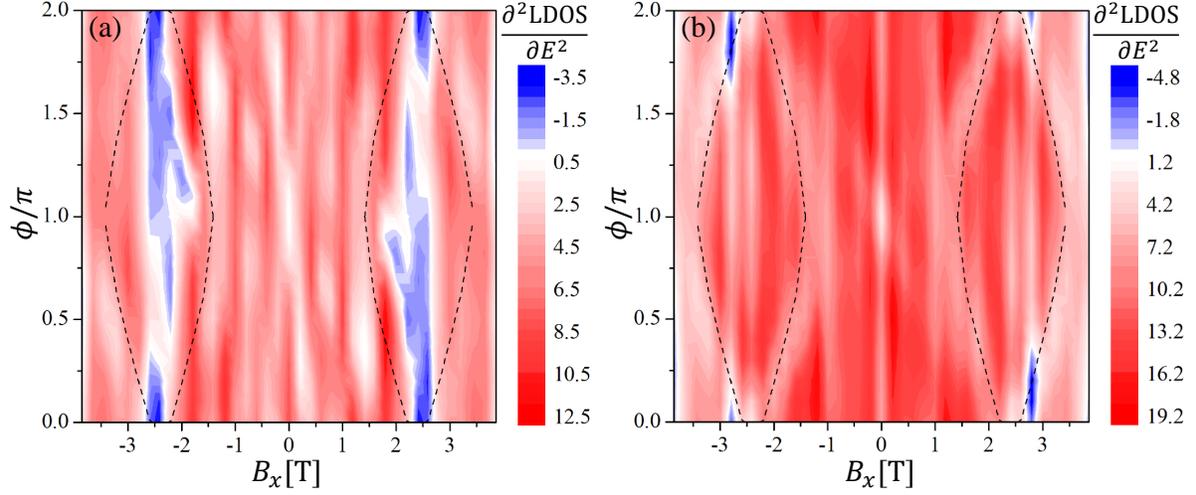

**Figure S20  Predicted zero-bias curvature for a 400-nm junction.** Calculated $\partial^2\mathrm{LDOS}/\partial E^2$ (in a.u.) at $E = 0$ for a Josephson junction of width $W = 400$ nm as a function of magnetic field $B_x$ and the phase difference $\phi$ between the superconducting regions. Panel (a) shows the LDOS computed from a $100\mathrm{nm} \times 100\mathrm{nm}$ area at the upper edge of the normal region (orange area in Figure S17). Panel (b) shows the LDOS computed from a $100\mathrm{nm} \times 100\mathrm{nm}$ area in the center of the normal region (yellow area in Figure S17). The dashed lines indicate the boundaries of the topological phase obtained from an infinitely long junction $L \to \infty$.

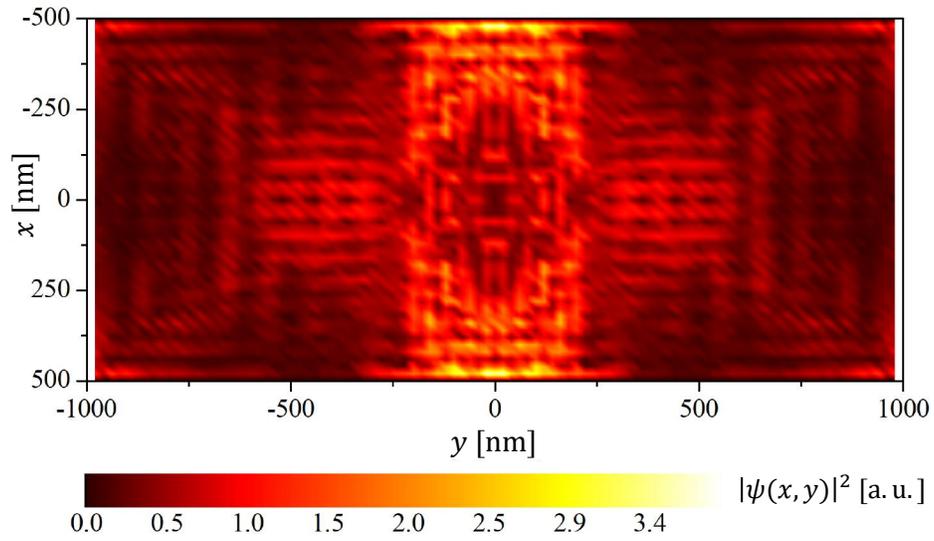

**Figure S21    Wavefunction of the low-energy state for a 400-nm junction.** Probability amplitudes $|\Psi(x,y)|^2$ of the low-energy state at $B_x = 2.4$ T and $\phi = 1.8\pi$.



 \end{document}